\documentclass{aa}  
\usepackage{graphicx,upgreek,txfonts,aas_macros}
\usepackage{hyperref}
\begin{document} 

\def\xmm {\emph{XMM--Newton}}
\def\cxo {\emph{Chandra}}
\def\swift {\emph{Swift}}
\def\frm {\emph{Fermi}}
\def\igr {\emph{INTEGRAL}}
\def\sax {\emph{BeppoSAX}}
\def\xte {\emph{RXTE}}
\def\ein {\emph{Einstein}}
\def\rst {\emph{ROSAT}}
\def\asca {\emph{ASCA}}
\def\hst {\emph{HST}}
\def\nst {\emph{NuSTAR}}
\def\src {\mbox{XMMU\,J122939.7+075333}}
\def\gc {\mbox{RZ\,2109}}
\def\flux {\mbox{erg cm$^{-2}$ s$^{-1}$}}
\def\lum {\mbox{erg s$^{-1}$}}
\def\nh {$N_{\rm H}$}

   \title{Recurrent X-ray flares of the black hole candidate in the globular cluster \gc\ in NGC\,4472}
\titlerunning{Recurrent X-ray flares of \src}
\authorrunning{Tiengo et al.}

   \author{A. Tiengo,\inst{1,2,3} P. Esposito,\inst{1,2} M. Toscani,\inst{4,5} G. Lodato,\inst{4} M. Arca Sedda,\inst{6} S. E. Motta,\inst{7} F. Contato,\inst{8} M. Marelli,\inst{2} \mbox{R. Salvaterra},\inst{2} A. De Luca\inst{2,3}
          }

   \institute{Scuola Universitaria Superiore IUSS Pavia, Palazzo del Broletto, piazza della Vittoria 15, I-27100 Pavia, Italy\\
   e-mail:
             \href{mailto:andrea.tiengo@iusspavia.it}{andrea.tiengo@iusspavia.it}
   \and Istituto Nazionale di Astrofisica, Istituto di Astrofisica Spaziale e Fisica Cosmica di Milano, 
      via A. Corti 12, I-20133 Milano, Italy
      \and Istituto Nazionale di Fisica Nucleare, Sezione di Pavia, via A. Bassi 6, I-27100 Pavia, Italy
      \and Dipartimento di Fisica "Aldo Pontremoli", Universit\`a degli Studi di Milano, via G. Celoria 16, I-20133 Milano, Italy
      \and Laboratoire des 2 Infinis - Toulouse (L2IT-IN2P3), Universit\'e de Toulouse, CNRS, UPS, F-31062 Toulouse Cedex 9, France
      \and Astronomisches Rechen-Institut. Zentrum f\"{u}r Astronomie der Universit\"{a}t Heidelberg, M\"onchhofstrasse 12-14, Heidelberg, D-69120, Germany
      \and Istituto Nazionale di Astrofisica, Osservatorio Astronomico di Brera, via E. Bianchi 46, I-23807 Merate (LC), Italy
      \and Department of Physics, University of Milano-Bicocca, Piazza della Scienza 3, I-20126 Milano, Italy
                }
  \date{Received 15 November 2021; accepted 15 February 2022}

  \abstract{
  We report on the systematic analysis of the X-ray observations of the ultra-luminous X-ray source \src\ located in the globular cluster RZ~2109 in the Virgo galaxy NGC\,4472. The inclusion of observations and time intervals ignored in previous works and the careful selection of extraction regions and energy bands have allowed us to identify new flaring episodes, in addition to the ones that made it one of the best black hole candidates in globular clusters. Although most observations are too short and sparse to recognize a regular pattern, the spacing of the three most recent X-ray flares is compatible with a $\approx$34 hours recurrence time. If confirmed by future observations, such behavior, together with the soft spectrum of the X-ray flares, would 
  be strikingly similar to the quasi-periodic eruptions recently discovered in galactic nuclei. 
  Following one of the possible interpretations of these systems and of a peculiar class of extra-galactic X-ray transients, we explore the possibility that \src\ might be powered by the partial disruption of a white dwarf by an intermediate mass ($M\approx700\,M_{\odot}$) black hole.}

  \maketitle

\section{Introduction}

\src\ is an ultra-luminous X-ray source (ULX) in the \gc\ globular cluster (GC) in the Virgo's giant elliptical galaxy NGC\,4472 (M49; \citealt{rhode01}), at a distance of 17.1 Mpc \citep{tully08}. Based on its luminosity greater than the Eddington limit for a neutron star, and on its variability (ruling out that its high luminosity might be due to the combination of multiple X-ray sources), \src\ is among the most robust black hole (BH) candidates in a GC \citep{maccarone07}. The census of BHs in GCs is an important route to test models of BH formation and of GC dynamical evolution (e.g., \citealt{morscher15,arcasedda18,giesers19}). 

\src\ is peculiar also in the optical band: its spectrum shows a very bright and broad [O\,\textsc{iii}] emission line, which implies a large overabundance of oxygen and the presence of high speed outflows \citep{zepf07}. Moreover, this line appears to be emitted by a spatially resolved nebula \citep{peacock12} and displays pronounced flux variability \citep{dage19}. The optical and X-ray peculiarities of \src\ have been interpreted by invoking different scenarios: a stellar mass BH accreting close to its Eddington limit from a low mass companion \citep{zepf08}, a nova shell illuminated by an external X-ray source \citep{ripamonti12}, 
the tidal disruption (or detonation; \citealt{irwin10})
of a white dwarf (WD; \citealt{clausen11}),
caused by an intermediate mass BH (IMBH). 

The X-ray variability of \src\ is a key piece of the puzzle to identify the nature of the compact object, the origin of the accreting matter and the physical processes producing its powerful electromagnetic emission. 
Marked variability has been reported both within single observations \citep{maccarone07,stiele19} and by comparing the average flux of different observations \citep{joseph15,dage18}. 
In this paper, we report on the discovery of more episodes of significant and possibly regular variability in archival \xmm\ and \cxo\ observations
of \src, including 
time periods affected by high particle background that were excluded in previous analyses. The peculiar variability pattern of \src\ was singled out within a broader search for flaring sources in the soft X-ray sky, carried out via tools developed as part of the EXTraS project \citep{deluca21extras}
that were applied to all the \xmm\ observations performed up to the end of 2018.
Thanks to the analysis of a larger data set, including also archival \ein\ and \rst\ observations, and a better characterization of high and low flux states, we were able to perform a more comprehensive and detailed study of the \src\ X-ray spectra at different flux levels and of its long term evolution.

\section{X-ray observations and data analysis}

\subsection{XMM--Newton observations}

\src\ was observed 8 times by \xmm\ from 2002 to 2016. The observation log is reported in Table~\ref{data}.
All observations were processed using \texttt{SAS 18.0.0}  and the most recent calibration files. Since the exposures of the three EPIC cameras are not always perfectly simultaneous and the EPIC-pn \citep{struder01} is much more sensitive than the EPIC-MOS \citep{turner01} in the soft energy band, where most of the variable X-ray emission of \src\ is concentrated, the analysis of the \xmm\ observations will be based mainly 

on EPIC-pn data.
The data of the two EPIC MOS cameras, when available, were checked for consistency.

All the EPIC cameras were equipped with the Thin filter and operated in Full Frame mode.
To subtract properly the background contribution, even during periods of high contamination by particle flares, we used a relatively small extraction region (a circle with a 15$^{\prime\prime}$ radius, containing $\sim$70\% of the counts from a point-source) for the source and a $\sim$10 times larger nearby region (on the same chip and free of apparent X-ray sources) to estimate and subtract the local background. According to standard procedures, only single- and double-pixel EPIC-pn events, with the PATTERN flag from 0 to 4, were selected.

\subsection{Chandra observations}
\cxo\ observed the position of \src\ 12 times between 2000 and 2019 with different CCDs of the ACIS \citep{garmire03} S and I arrays (Table\,\ref{data}). The data were processed and reduced using the \texttt{CIAO 4.11} software package and the calibration files in \texttt{CALDB 4.8.2}. The source data were selected from elliptical regions of sizes chosen so as to include $\sim$95\% of the PSF (depending on the off-axis distance, their semi-major axes vary from $\sim$7.5 to 13.5 arcsec). The background was estimated from nearby regions. The spectra,  the spectral redistribution matrices and the ancillary response files were created using \texttt{specextract}. For the observations in which \src\ was not detected, we set upper limits using the \texttt{CIAO} tool \texttt{aprates}.

\subsection{ROSAT observations}

The {\it Roentgen  Satellite} (\rst; \citealt{pfeffermann87}) pointed at the NGC~4472 galaxy in December 1992 (a short 7 ks exposure with the HRI and then a deeper 25 ks with the PSPC) and in 1994, between June 19 and July 9 (27 ks with the HRI). The average off-axis angles and source net count rates are reported in Table~\ref{data}.
The HRI rates are taken from the \rst\ HRI Pointed Observations catalog (1RXH; \citealt{1RXH}), whereas the PSPC rate was computed by extracting the source counts from a circle with a 54$^{\prime\prime}$ radius and subtracting the background contribution from a nearby region. The same regions were used also to extract the source light-curve, spectrum and corresponding spectral matrix. The HRI light-curves, instead, were extracted from 10$^{\prime\prime}$ circular regions.

\subsection{Einstein observations}

The \ein\ observatory \citep{giacconi79} observed the Virgo field including NGC~4472 multiple times between the end of 1979 and the beginning of 1981. In particular, the \ein\ High Resolution Imager (HRI) instrument observed \src\ at an off-axis angle of $\sim$7$^{\prime}$ for a cumulative net exposure time of $\sim$33 ks (Table~\ref{data}). The analysis reported here is based on the image and automatic source detection produced by the \ein\ HRI Catalog of ESTEC Sources (HRIEXO).\footnote{https://heasarc.gsfc.nasa.gov/W3Browse/einstein/hriexo.html} The image distinctly displays a point source at the position of \src\ and the catalog reports a source with $\sim$80 net counts at a distance $<$6$^{\prime\prime}$ from its coordinates. 

\begin{table*}
\caption{
X-ray observations of \src.\label{data}}
\centering                          
\begin{tabular}{llccccc}
\hline\hline    
Instrument & Obs.ID & Start date & End date & Exposure & Off-axis& Net rate\\   
& & \multicolumn{2}{c}{YYYY-MM-DD hh-mm-ss} & (ks) & (arcmin) & (cts/ks)\\
\hline   
Einstein/HRI & 7068 & 1979-12-31 11:10:17 & 1981-01-10 13:47:58 & 32.8 & 6.8 & $2.5\pm0.4^a$\\
ROSAT/HRI & 600216 & 1992-12-10 22:34:54 & 1992-12-14 11:37:52 & 7.0 & 7.8 & $5.3\pm1.1^b$\\
ROSAT/PSPC & 600248 & 1992-12-26 23:04:44 & 1992-12-29 20:21:42 & 25.1 & 6.8 & $17.0\pm1.1^c$\\
ROSAT/HRI & 600216 & 1994-06-19 00:22:08 & 1994-07-09 03:15:59 & 27.1 & 5.9 & $5.3\pm0.5^b$\\
CXO/ACIS-I & 322 & 2000-03-19 12:01:38 & 2000-03-19 15:42:42 & 10.4 & 6.9& $4.2\pm0.6^d$\\
CXO/ACIS-S & 321 & 2000-06-12 01:48:50 & 2000-06-12 13:31:49 & 39.6 &6.9& $8.8\pm0.5^d$\\
XMM/PN & 0112550601 & 2002-06-05 13:20:09 & 2002-06-05 16:48:32 & 11.1 & 6.4& 49.6$\pm$2.2$^e$\\
XMM/PN & 0200130101 & 2004-01-01 03:46:56 & 2004-01-02 05:19:03 & 79.0 & 6.7& 10.1$\pm$0.5$^e$ \\
XMM/PN & 0510011501 & 2008-01-07 04:16:44 & 2008-01-07 04:50:22 & 1.1 & 0 & 45.5$\pm$8.7$^e$ \\
CXO/ACIS-S & 8095 & 2008-02-23 20:10:45 & 2008-02-23 21:53:42 & 5.1 &6.5& $10.0\pm1.4^d$\\
CXO/ACIS-S & 11274 & 2010-02-27 18:16:16 & 2010-02-28 05:58:04 & 39.7 & 6.7& $<$0.4$^d$\\
CXO/ACIS-S & 12978 & 2010-11-20 10:14:49 & 2010-11-20 16:17:55 & 19.8 &0& $0.8\pm0.2^d$\\
CXO/ACIS-S & 12889 & 2011-02-14 12:08:04 & 2011-02-16 02:48:40 & 135.6 & 6.7& $2.1\pm0.1^d$\\
CXO/ACIS-S & 12888 & 2011-02-21 20:24:22 & 2011-02-23 17:49:49 & 159.3 &6.7& $1.0\pm0.1^d$\\  
XMM/PN & 0722670301 & 2014-01-09 19:05:24 & 2014-01-10 08:09:17 & 13.8 & 14.9 &16.1$\pm$1.3$^e$\\
XMM/PN & 0722670601 & 2014-01-13 12:09:00 & 2014-01-13 20:13:02 & 25.8 & 14.9& 3.1$\pm$0.5$^e$ \\
CXO/ACIS-I & 15760 & 2014-04-26 03:27:40 & 2014-04-26 12:04:26 & 29.4 & 9.3&$<$0.5$^d$\\
CXO/ACIS-S & 16260 & 2014-08-04 18:47:09 & 2014-08-05 02:24:31 & 24.7 & 2.0 &$2.3\pm0.4^d$\\
CXO/ACIS-S & 16261 & 2015-02-24 09:10:20 & 2015-02-24 16:08:51 & 22.8 & 2.0&$<$0.7$^d$ \\
XMM/PN & 0761630101 & 2016-01-05 09:05:51 & 2016-01-06 17:03:12 & 100.7 & 0& 17.4$\pm$0.5$^e$\\
XMM/PN & 0761630201 & 2016-01-07 10:09:24 & 2016-01-08 17:22:51 & 98.3 & 0& 11.5$\pm$0.5$^e$ \\
XMM/PN &0761630301 & 2016-01-09 08:48:39 & 2016-01-10 16:29:43 & 100.1 & 0 & 7.6$\pm$0.4$^e$ \\
CXO/ACIS-S & 16262 & 2016-04-30 00:54:33 & 2016-04-30 08:32:53 & 24.7 & 1.0& $0.50\pm0.15^d$\\
CXO/ACIS-S & 21647 & 2019-04-17 03:39:21 & 2019-04-17 12:35:16 & 29.7 & 1.0&$<$0.7$^d$ \\

\hline                                   
\end{tabular}
\tablefoot{$^a$From the HRIEXO catalog. 
$^b$From the 1RXH catalog. 
$^c$In the 0.15--2 keV energy band, within 54$^{\prime\prime}$. 
$^d$In the 0.3--2 keV energy band; we used elliptical regions including $\sim$95\% of the PSF (their semi-major axes go from $\sim$7.5 to 13.5 arcsec, depending on the off-axis). The upper limits are at a 3$\sigma$ confidence level.
$^e$In the 0.15--2 keV energy band, within 15$^{\prime\prime}$.}
\end{table*}

\section{Results}

\subsection{Short-term X-ray variability}\label{shortVar}

\begin{figure}
\centering
\resizebox{\hsize}{!}{\includegraphics[angle=0]{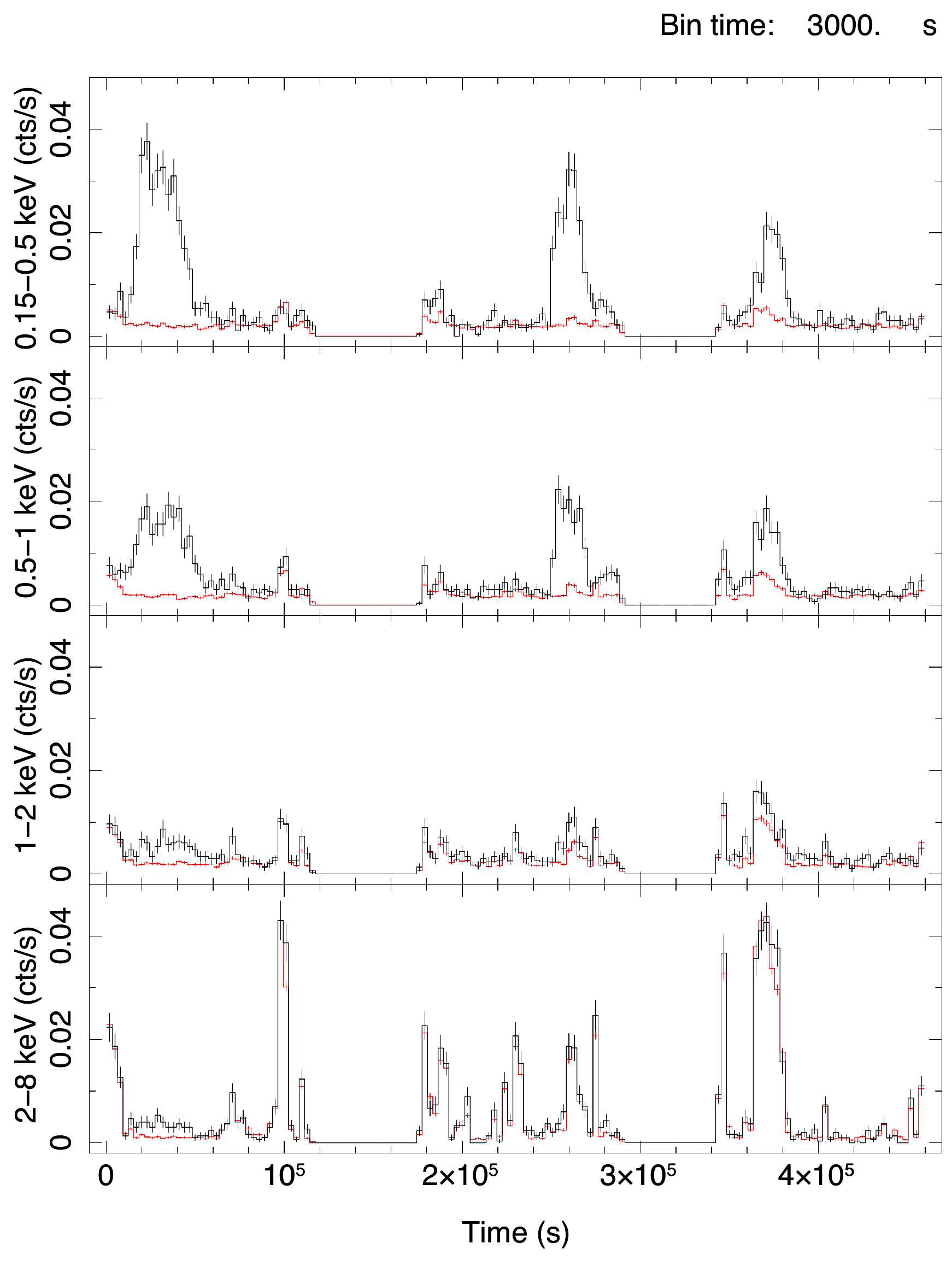}}
\caption{EPIC-pn light curves (from top to bottom: 0.15--0.5 keV, 0.5--1 keV, 1--2 keV, 2--8 keV) extracted from the source (black) and background (red) regions during the three \xmm\ observations performed in 2016 (0761630101, 0761630201 and 0761630301). The background count-rates (and error bars) are rescaled by a factor 10 to account for the different size of the source and background regions.
 \label{PN2016lc}}
\end{figure}

The light curves of \src\ in four energy bands registered by the EPIC-pn during three consecutive \xmm\ orbits in January 2016 are shown in Figure~\ref{PN2016lc}. Three flares are clearly detected, mainly in the soft energy bands ($E<1$\,keV). The first two flares were already discovered and discussed by \citet{stiele19}, whereas the one occurring during the last observation was missed due the rejection of time intervals with high particle background. As shown mainly by the bottom panel of Figure~\ref{PN2016lc}, a background flare occurs at the approximate time of the source flare, but 
its intensity in the softest energy bands (upper panels of Figure~\ref{PN2016lc}) is much smaller than the flux increase of the point-source.

Although the profile of the three peaks significantly deviates from a Gaussian, we estimate their time separations and widths with a simple model consisting of a constant plus three Gaussian curves. Applying this model to the 0.15--2\,keV background-subtracted EPIC-pn light-curve ($\chi^2_{\rm red}=1.6$ for 107 degrees of freedom, d.o.f.), the distance between the centroids of the first two Gaussian peaks and between the last two are $229\pm0.5$  and $111.9\pm0.7$\,ks, respectively. The FWHM of the three Gaussian peaks are $26.6\pm0.9$, $14.1\pm0.7$, and $16.7\pm1.4$\,ks. 

\begin{figure}[h!]
\centering
\resizebox{\hsize}{!}{\includegraphics[angle=0]{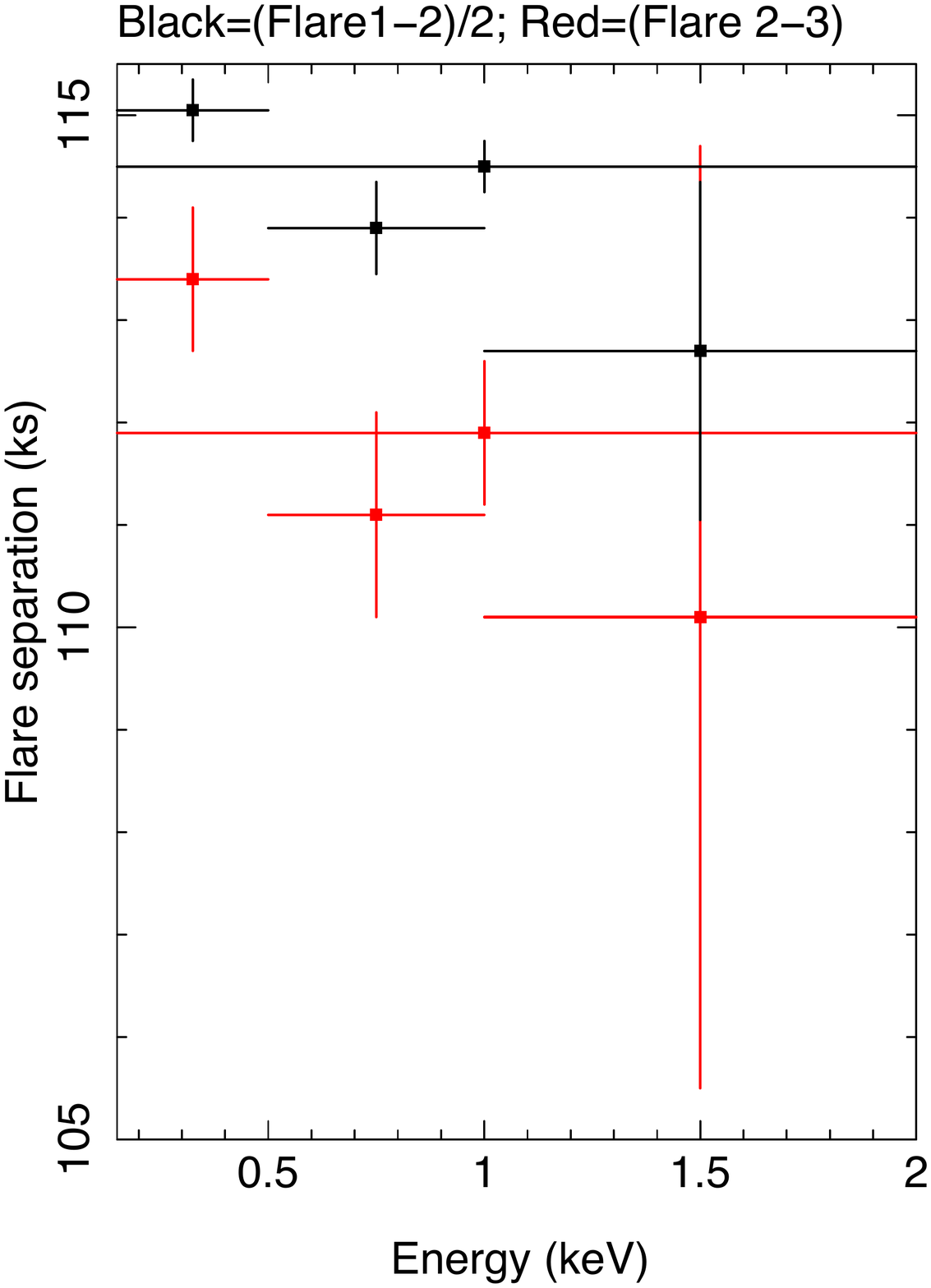}\includegraphics[angle=0]{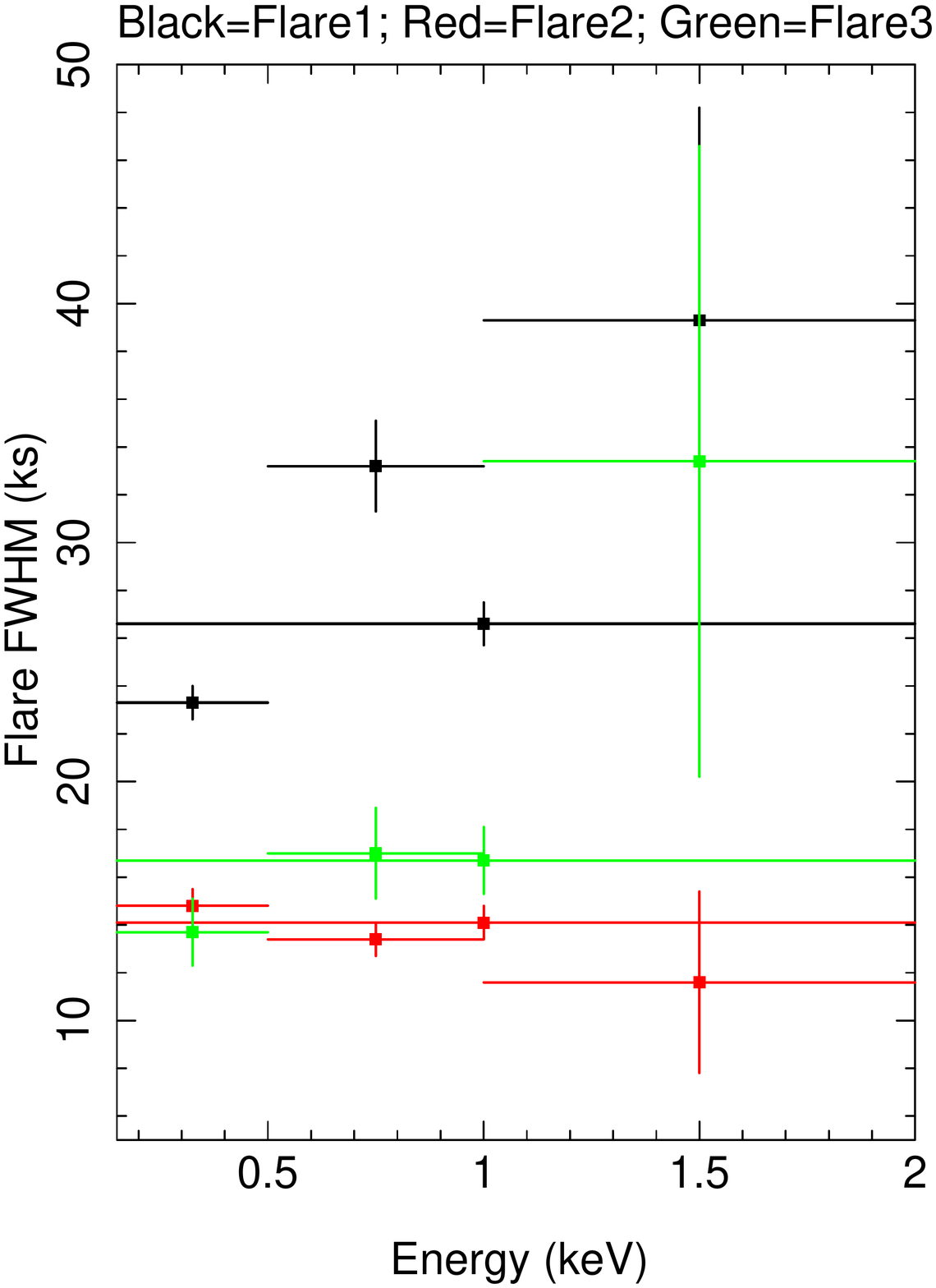}}
\caption{Energy dependence of the peak separation (left panel) and FWHM (right panel) of the three flares observed in the 2016 \xmm\ observations, in three energy bands and in the 0.15--2 keV total range. The separation between the first two flares has been divided by 2, in order to compare it with the separation between the last two flares.
 \label{LCpeaks}}
\end{figure}

We have also checked the energy dependence of these parameters by fitting with the same model also the background-subtracted EPIC-pn light-curves in the 0.15--0.5, 0.5--1, and 1--2\,keV energy bands. Although the left panel of Figure\,\ref{LCpeaks} shows a
slight decrease of the peak separation with energy, this trend is not statistically significant. 
The FWHM (right panel of Figure\,\ref{LCpeaks}) displays a significant increase with energy for the first peak 
whereas a statistically significant trend cannot be derived for the second 
and third 
peaks.

It is interesting to note that the separation between the first two flares approximately doubles the distance between the last two peaks and that, in case of a fixed recurrence time, one flare could have been missed due to its occurrence in the middle of the $>$60\,ks gap between the first two observations. A possible indication of such a flare might be present in the PN light-curve in the softest energy band (top panel of Figure\,\ref{PN2016lc}), where a small count-rate excess emerges at the beginning of the second exposure. A fit of the first 60 ks of the background-subtracted light-curve  with a constant function provides $\chi^2_{\rm red}$=2.4 for 19 d.o.f., corresponding to a null-hypothesis probability of 5$\times$10$^{-4}$. A similar excess is also present in the MOS light-curve in the 0.15--1 keV energy band, but with a lower 
statistical significance 
(4$\times$10$^{-3}$), as expected due to the much smaller effective area of the MOS cameras at very low energy, 
which is only partly compensated by the slightly earlier start of the MOS exposures.

\begin{figure}
\centering
\resizebox{\hsize}{!}{\includegraphics[angle=0]{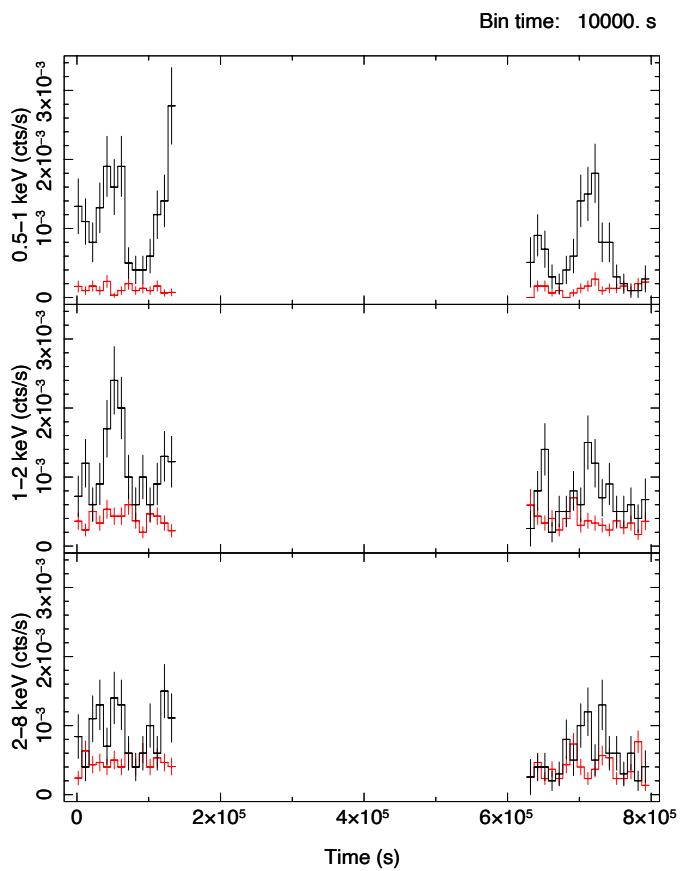}}
\caption{ACIS-S light curves (from top to bottom: 0.5--1 keV, 1--2 keV, 2--8 keV) extracted from the source (black) and background (red) regions during the two \cxo\ observations performed in 2011 (12888 and 12889). The background count-rates (and error bars) are rescaled by a factor 3 to account for the different size of the source and background regions. \label{CXO2011lc}}
\end{figure}

The only other archival observations of \src\ that allow us to test the hypothesis of a possible flare recurrence on time-scales $>$1\,day are the two long observations performed by \cxo\ in 2011 at a distance of one week (Table~\ref{data}).
The ACIS-S light-curves in three energy bands\footnote{Due to the reduced effective area of ACIS-S below 0.5\,keV, too few counts were collected to produce a meaningful light-curve in the 0.15--0.5 keV energy band.} are shown in Figure\,\ref{CXO2011lc}: significant variability is clearly detected in both observations,\footnote{This result is in contrast with \citet{joseph15}, where only observation 12888 was reported to be significantly variable, based on the fit of the light-curves with a constant function. If we fit with a constant the 0.5--2 keV background-subtracted light-curve of each observation, binned at 10 ks so that the minimum number of counts per bin permits the application of the $\chi^2$ statistic,  we obtain unacceptable $\chi^2_{\rm red}$ values: 4.5 for 13 d.o.f. and 3.2 for 16 d.o.f. for observation 12889 and 12888, respectively.} against a fairly stable background.

Fitting the 0.5--2 keV background-subtracted light-curve of each observation with a constant plus two Gaussian curves ($\chi^2_{\rm red}=1.5$ for 7 d.o.f. and $\chi^2_{\rm red}=1.2$ for 10 d.o.f., respectively), the distance between the two peaks in the first and second observation are $78.8\pm5.3$ and $68.4\pm3.6$\,ks, respectively. The first and the last peaks, whose position is rather well defined, occur at a distance of $668.7\pm3.6$\,ks. We note that such distance would correspond to 9 cycles of a 73.4\,ks periodicity, which would be compatible with the separations of the peaks detected in the single observations.
On the other hand, only these peaks are clearly observed in multiple energy bands and are fully included in the observations and so, considering also the structured light-curve profile, we cannot exclude a longer recurrence time. In particular, the separation between them would be compatible with 6 cycles of a $\sim$112\,ks period, consistent with the recurrence time suggested by the 2016 observations.

The FWHM of the four peaks are $24.3\pm4.9$, $25.2\pm9.4$, $12\pm7.1$, and $51.1\pm10.4$\,ks. They are consistent with the range of peak widths 
measured in the EPIC-pn light-curves (right panel of Figure~\ref{LCpeaks}).

In addition to the prominent variability of \src\ during these long observations and in the 2004 \xmm\ observation already reported by \citet{maccarone07}, we detected significant changes in the source count-rate also in the 2008 \xmm\ observation and in two out of the four observations performed in 2014 (Table~\ref{data}). 

\begin{figure}
\centering
\resizebox{\hsize}{!}{\includegraphics[angle=0]{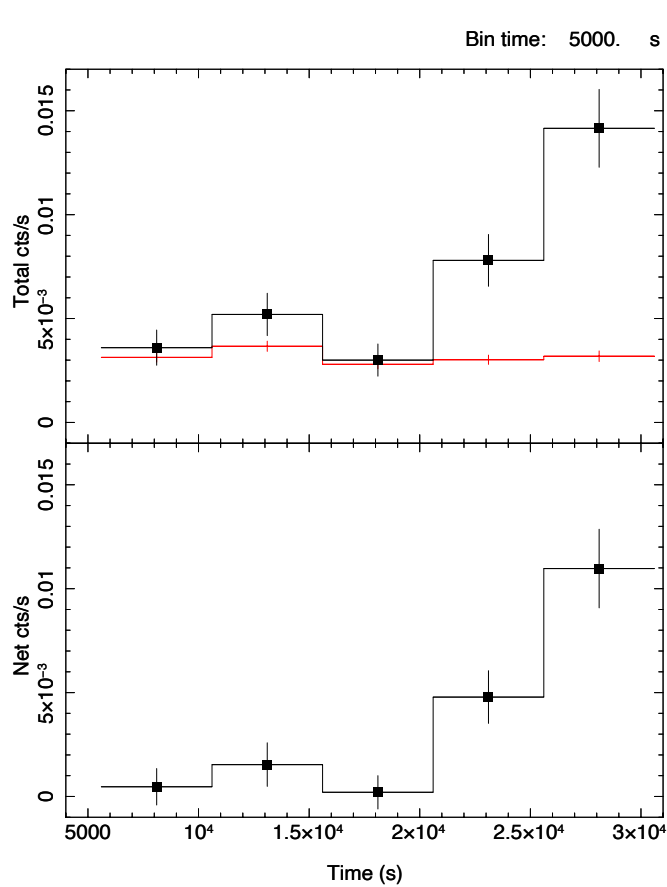}}
\caption{Top panel: EPIC-pn light-curve, in the 0.15--2 keV energy band, extracted from the source (black squares) and background (red) regions during observation 0722670601. The background count-rates (and error bars) are rescaled by a factor 12 to account for the different size of the source and background regions. Bottom panel: Background-subtracted EPIC-pn light curve, in the same energy band.
 \label{PN2014lc}}
\end{figure}

In January 2014, \src\ was observed twice by \xmm\ at such a large off-axis angle that it could be detected only by the EPIC-pn, because it was outside the EPIC-MOS field of view. 
No 
source variability was detected in the first observation (0722670301), whereas
during the second exposure (0722670601),  
we detected a significant count-rate increase, without any corresponding background variation (Figure~\ref{PN2014lc}). If we fit the 0.15--2 keV background-subtracted EPIC-pn light-curve with a constant function, we get an unacceptable $\chi^2_{\rm red}$ of 8.8 for 4 d.o.f..

\begin{figure}
\centering
\resizebox{\hsize}{!}{\includegraphics[angle=0]{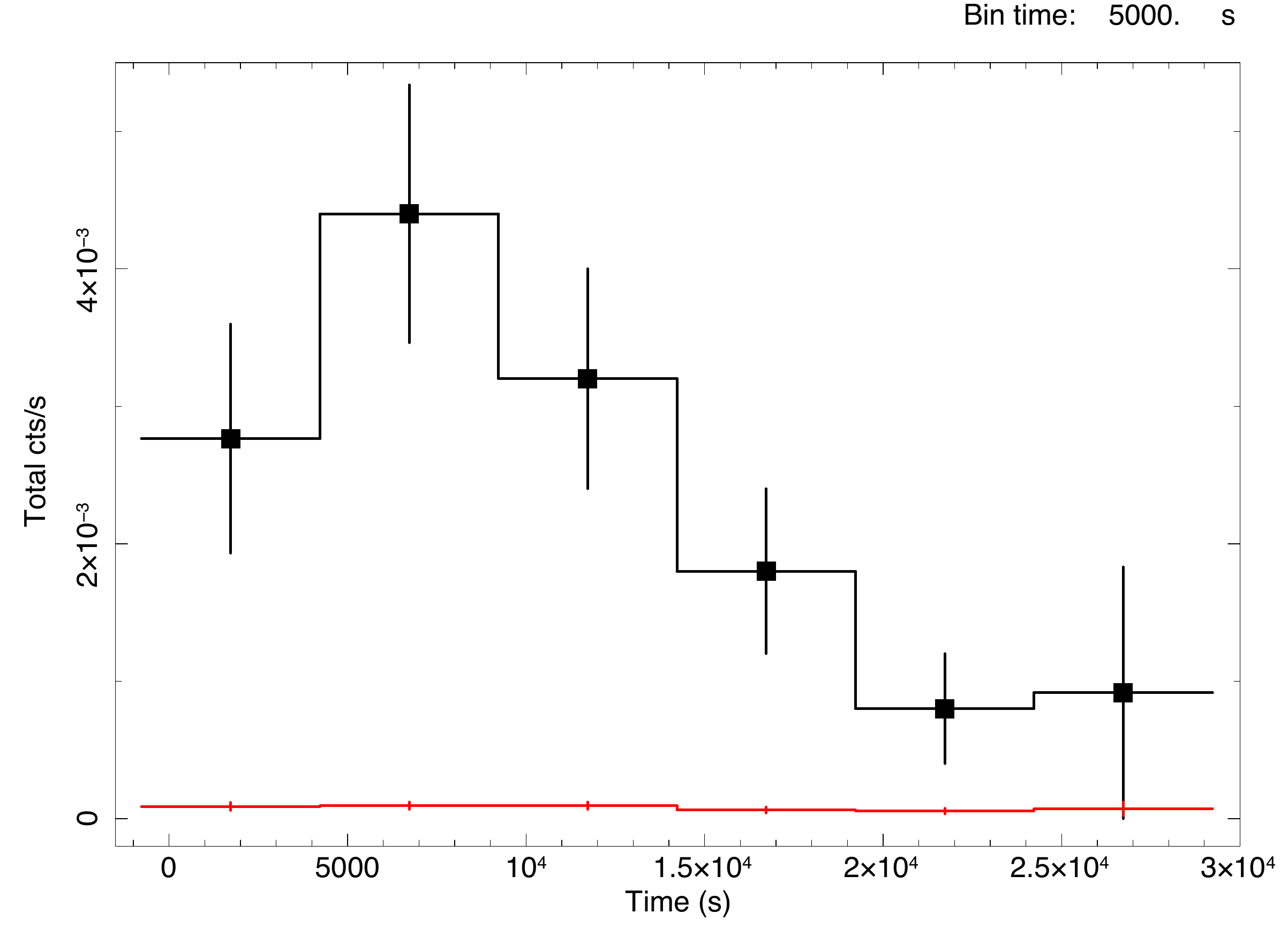}}
\caption{ACIS-S light-curve, in the 0.3--2 keV energy band, extracted from the source (black squares) and background (red) regions during observation 16260. The background count-rates (and error bars) are rescaled by a factor 25 to account for the different size of the source and background regions. 
 \label{CXO2014lc}}
\end{figure}

\cxo\ observed \src\ three times between April 2014 and February 2015, but detected it significantly only in one occasion (observation 16260). The corresponding light-curve in the 0.3--2 keV energy band is shown in Figure~\ref{CXO2014lc} and appears to be variable. By fitting the background-subtracted light-curve with a constant count-rate we obtain a $\chi^2_{\rm red}$=3.7 for 5 d.o.f., corresponding to a null-hypothesis probability of 2$\times$10$^{-3}$. 

\begin{figure}
\centering
\resizebox{\hsize}{!}{\includegraphics[angle=0]{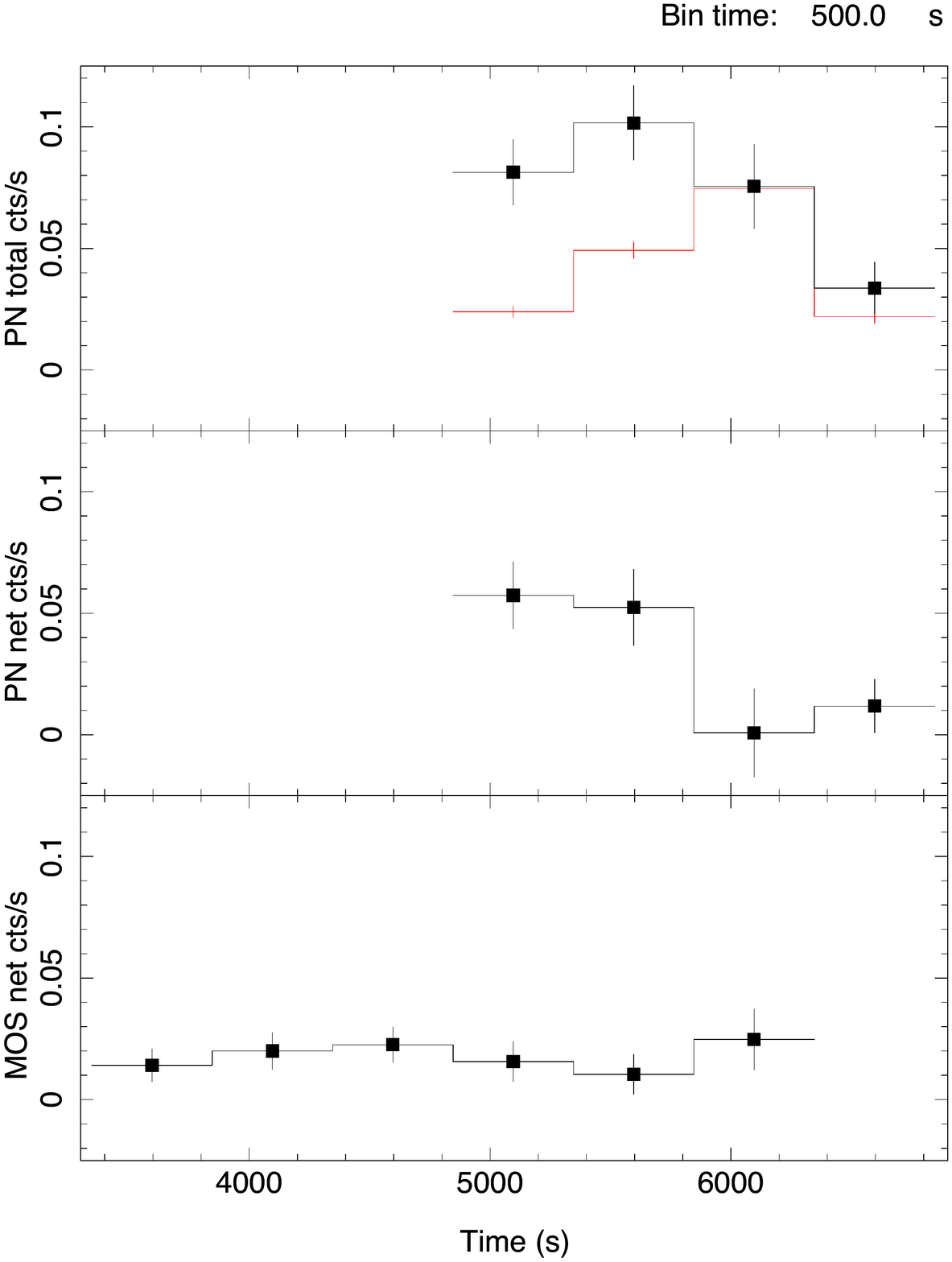}}
\caption{Top panel: EPIC-pn light curve in the 0.15--2 keV energy band extracted from the source (black squares) and background (red) regions during the 2008 \xmm\ observation (0510011501). The background count-rates (and error bars) are rescaled by a factor 10 to account for the different size of the source and background regions. Middle panel: EPIC-pn background-subtracted light-curve, in the same energy band. Bottom panel: EPIC MOS1+MOS2 background-subtracted light-curve, in the same energy band.
 \label{PNMOS2008lc}}
\end{figure}

Finally, during a very short observation performed in 2008 and affected by strongly variable background, the EPIC-pn detected a sharp decrease in the net count-rate of \src\ (Figure~\ref{PNMOS2008lc}). 
Since the statistical significance of the deviation from a constant rate is rather low ($\chi^2_{\rm red}$=3.8 for 3 d.o.f., corresponding to a null-hypothesis probability of 10$^{-2}$),
we checked also the light-curve accumulated by the EPIC MOS1 and MOS2 detectors, which started observing earlier. The background-subtracted light-curve obtained combining the data of the two EPIC-MOS cameras is perfectly consistent with a constant count rate ($\chi^2_{\rm red}$=0.4 for 5 d.o.f.; bottom panel of Figure~\ref{PNMOS2008lc}). However, we note that the EPIC-MOS observed \src\ mostly before the flux drop indicated by the EPIC-pn light-curve and it detected an average count-rate of (1.8$\pm$0.3)$\times$10$^{-2}$ cts\,s$^{-1}$, much larger than that observed during quiescent periods. For example, during the second 2016 observation (0761630201), the average EPIC-MOS count-rate was (0.39$\pm$0.03)$\times$10$^{-2}$ cts\,s$^{-1}$. If we separate the quiescent and flare time intervals by applying 
a count-rate threshold of 0.01 cts\,s$^{-1}$ to the 0.15--1 keV EPIC-pn light-curve, the count-rates during quiescence and the flare were (0.14$\pm$0.03)$\times$10$^{-2}$ cts\,s$^{-1}$ and (1.25$\pm$0.08)$\times$10$^{-2}$ cts\,s$^{-1}$, respectively.
Despite the much smaller efficiency of the EPIC-MOS at low energies and the earlier stop of the 2008 MOS exposures, these data 
support the marginal detection of a flare in the 2008 EPIC-pn data.

The light-curves extracted from the \rst\ data do not display any significant variability. However, since the satellite was in low Earth orbit, these light-curves are not continuous and they are less sensitive to variability on timescales of a few hours.

\subsection{Spectral analysis of the 2016 \xmm\ observations}\label{PN2016spec}

Thanks to the clear identification of three flares in the 2016 \xmm\ observations, the X-ray spectra of \src\ at different flux levels
could be compared better than in previous works (e.g., \citealt{maccarone07,joseph15,stiele19}). Specifically, we analysed the EPIC-pn spectra of the quiescent and flare states during each of the three observations adopting the same count-rate threshold as in Section~3.1.
The spectral analysis was performed using XSPEC v12.10.1. The absorption component was fixed at the foreground Galactic value (\nh=1.55$\times$10$^{20}$ cm$^{-2}$, assuming solar abundances from \citealt{wilms00}). A distance of 17.1 Mpc \citep{tully08} was assumed to compute the source luminosity (and derived quantities).

\begin{figure}[ht]
\begin{center}
\vspace{-0.8cm}
\hbox{
\includegraphics[height=7.5cm]{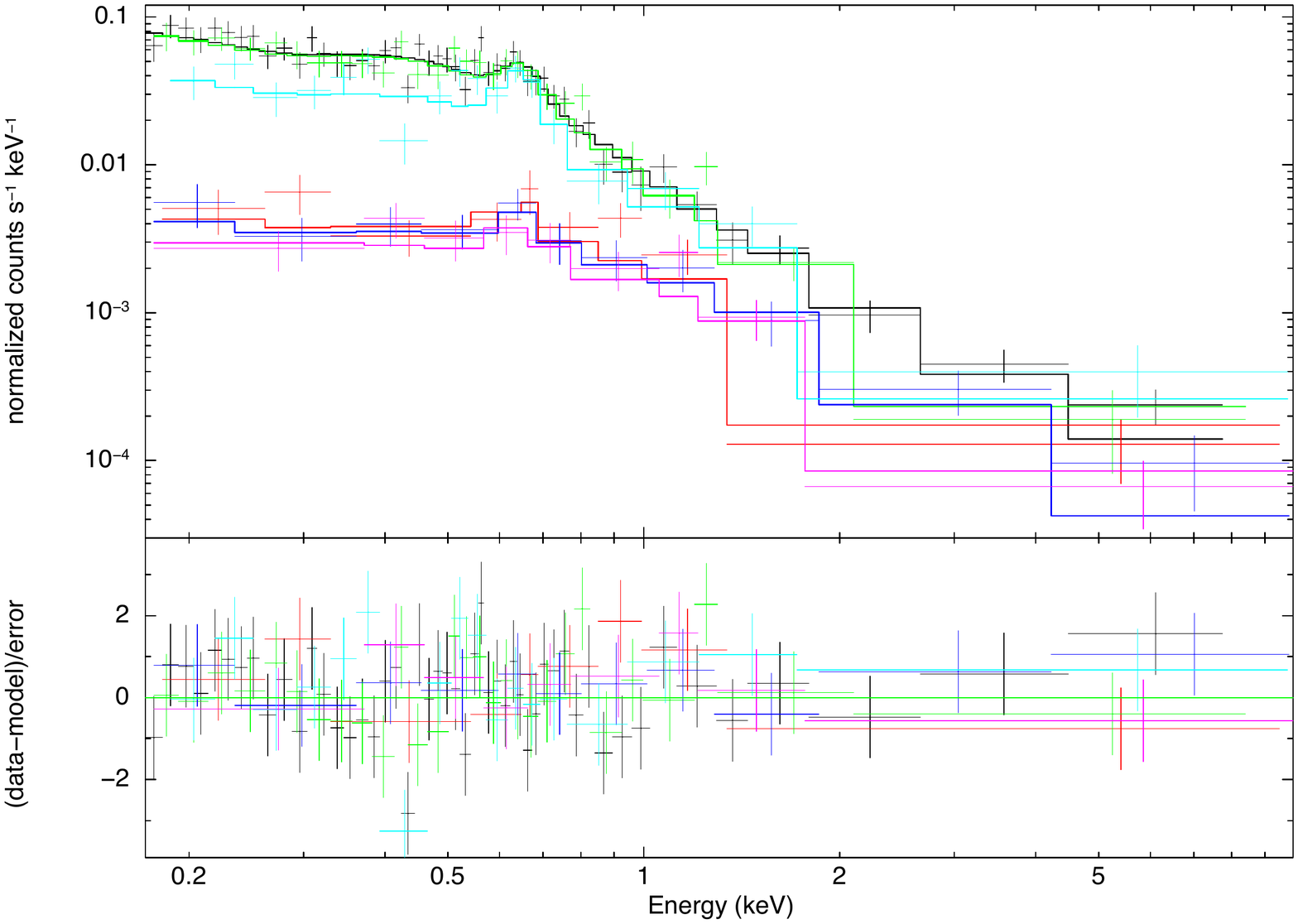}
\hspace{-1.0cm}
}
\end{center}
\vspace{-1cm}
\caption{\footnotesize
{EPIC-pn spectra (upper panel) and residuals with respect to the best-fit model with parameters in Table~\ref{SpecPar} (lower panel) of \src\ in observations 0761630101 (black during flare, red in quiescence), 0761630201 (green during flare, blue in quiescence) and 0761630301 (cyan during flare, purple in quiescence). }}
\label{spec}
\end{figure}

From a simultaneous fit of the six EPIC-pn spectra with a {\tt TBabs*(pegpwrlw + diskbb + gaussian)} model,
with the Gaussian line intrinsic width and centroid fixed at 0 and 0.65 keV, respectively, we obtain the parameters reported in Table~\ref{SpecPar}. 
The spectra and residuals with respect to the model are shown in Figure~\ref{spec}. 
As already noted in previous works, the spectrum during the flare is much softer. In this spectral deconvolution, the spectral softening can be attributed to the appearance of a disk component only during the flare and a corresponding increase, by about an order of magnitude, of the flux of the emission line, which, as already proposed by \citet{joseph15}, can be readily identified as O\,\textsc{viii}. At the same time, also the hard power-law component characterizing the bulk of quiescent emission significantly increases during flares, but only by a factor $\sim$2. Except for the normalization of the disk component of the last flare, which 
shows a lower peak flux in the soft energy band (Figure~\ref{PN2016lc}),
the parameters of the two spectral states are consistent across the three observations.

\begin{table*}
\caption{Best-fit parameters of the EPIC-pn spectra during flares and quiescence in the 2016 \xmm\ observations.\label{SpecPar}}
\centering                          
\begin{tabular}{lcccccc}
\hline\hline    
Model parameters$^a$ & obs1F$^b$ & obs1Q$^b$ & obs2F$^b$ & obs2Q$^b$ & obs3F$^b$ & obs3Q$^b$ \\   
\hline   
$\Gamma$ & 1.9$^{+0.1}_{-0.2}$ & 1.9$^c$ & 1.9$^c$ & 1.9$^c$ & 1.9$^c$ & 1.9$^c$ \\
PL flux$^d$ (10$^{-14}$ \flux) & 3.7$\pm$0.4 & 1.5$\pm$0.3 & 3.3$\pm$0.7 & 1.4$\pm$0.2 & 3.3$\pm$0.9 &1.1$\pm$0.2\\
T$_{\rm in}$ (eV)& 143$\pm$7 & 143$^c$ & 143$^c$ & 143$^c$ & 143$^c$ & 143$^c$ \\
R$_{\rm in}^e$ (km) & 6400$^{+800}_{-600}$ & $<$1100 & 6300$^{+800}_{-600}$ & $<$1100 & 4200$\pm$600 &$<$900\\
O\,\textsc{viii} norm.$^f$ (10$^{-6}$ photons cm$^{-2}$ s$^{-1}$) & 4.2$\pm$1.0 & 0.6$\pm$0.4 & 4.3$\pm$1.2 & 0.5$\pm$0.3 & 5.7$\pm$1.3 &0.4$\pm$0.3\\
\hline
\end{tabular}\\
\tablefoot{$^a$The model is {\tt TBabs*(pegpwrlw + diskbb + gaussian)} with absorption fixed to \nh=1.55$\times$10$^{20}$ cm$^{-2}$ and the emission line centroid and width fixed to 0.65 keV  and 0, respectively. All uncertainties are at 1 $\sigma$ and $\chi^2_{\rm red}$=1.06 for 267 d.o.f. $^b$F and Q indicate flaring and quiescent time intervals, respectively, during observations 0761630101 (obs1), 0761630201 (obs2) and 0761630301 (obs3).
$^c$Linked parameter. $^d$Unabsorbed flux of the power-law component in the 0.3--10 keV energy band. $^e$Apparent inner radius of the {\tt diskbb} model, assuming 
a disk inclination of 45$^{\circ}$. $^f$Flux of
Gaussian emission line with E=0.65 keV and $\sigma$=0.
}\\
\vspace{0.2cm}
\end{table*}

 Since in previous works (e.g., \citealt{maccarone07,joseph15}) it was proposed that the low flux states of \src\ might be caused by an excess of absorption, we tried to fit the six spectra keeping linked all the spectral parameters
 and adding a free additional absorption component, but the resulting fit was unacceptable
 ($\chi^2_{\rm red}$=1.38 for 276 d.o.f.; null hypothesis probability=2.4$\times$10$^{-5}$).

A much better fit ($\chi^2_{\rm red}$=1.13 for 271 d.o.f.; null hypothesis probability=6.5$\times$10$^{-2}$) is instead obtained by adding to the model of each spectrum also an overall free normalization factor. In this case, the three flare spectra are consistent with no additional absorption, and the quiescent spectra with an excess absorption of \nh$\sim$10$^{21}$ cm$^{-2}$. The overall normalization factors indicate that the flux of the third flare was $\sim$30\% lower than that of the previous ones and that during the quiescent states, in addition to the excess absorption, the intrinsic source flux decreased by a factor $\sim$4.

\subsection{Long-term X-ray variability}

\begin{figure}[ht]
\begin{center}
\hbox{
\resizebox{\hsize}{!}{\includegraphics{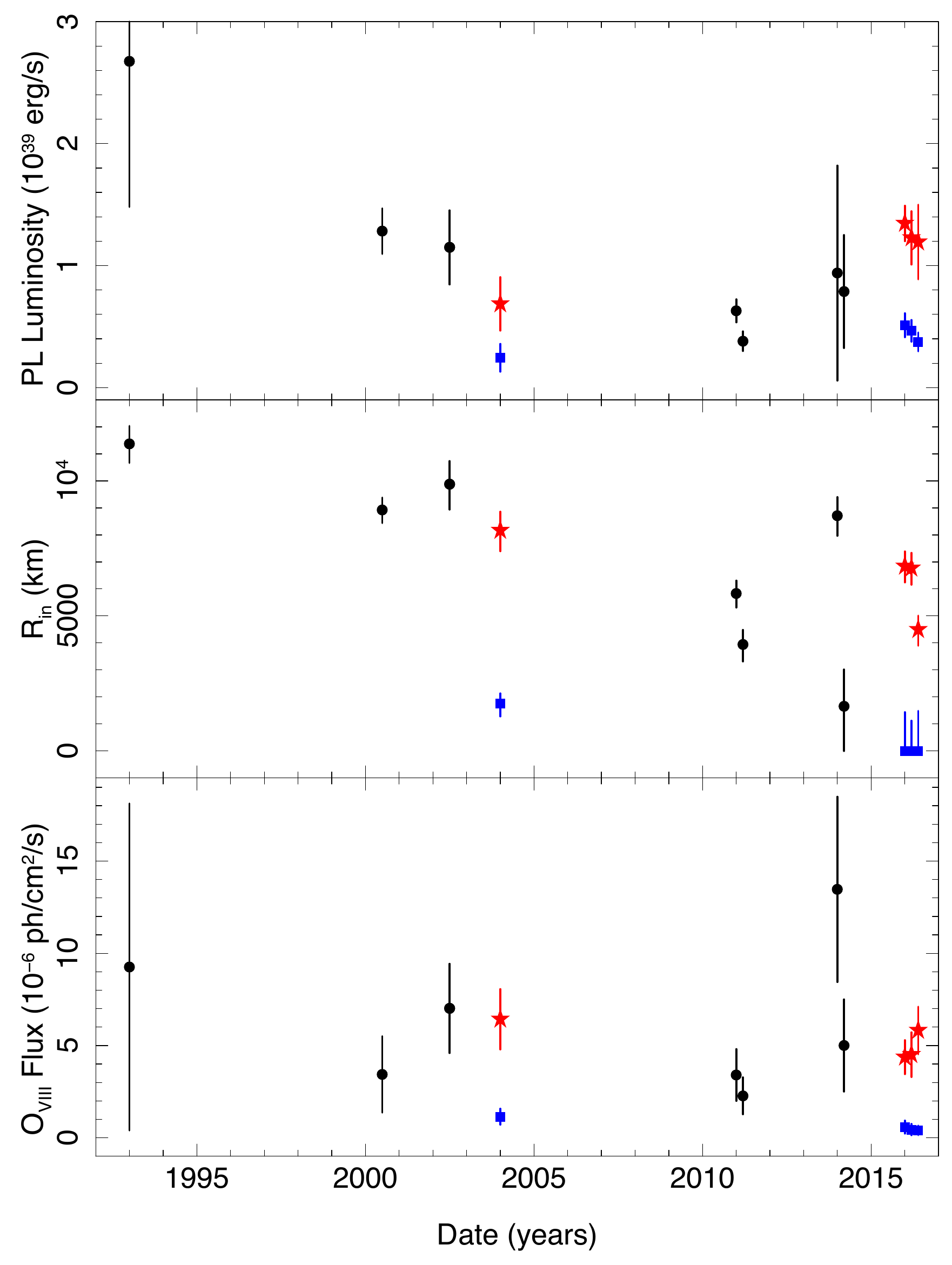}}
}
\end{center}
\vspace{-0.8cm}
\caption{\footnotesize
{Time evolution of the power-law 0.3--10 keV unabsorbed luminosity (upper panel), apparent inner disk radius (middle panel; from the diskbb model in XSPEC, assuming a disk inclination of 45$^{\circ}$) and Gaussian line flux (lower panel) obtained by fitting the available spectra with the same model as in Table~\ref{SpecPar}. Spectra extracted from the full observation, flare and quiescent time intervals are represented by black circles, red stars and blue squares, respectively. The time separations between the exposures performed during the same year have been enhanced for clarity.}}
\label{SpecParPlot}
\end{figure}

We applied the best-fit spectral model used for the 2016 observations, with only the normalizations of the three spectral components as free parameters, to all the available PSPC, EPIC-pn and ACIS-S spectra with sufficient count statistics, separating, when possible, the flare and quiescent time periods.\footnote{In addition to the spectra described in the previous section, we extracted the spectra from two different time intervals from the \xmm\ observation performed in 2004 \citep{maccarone07}.} This simple model provides acceptable fits to all the spectra, with the parameters displayed in Figure~\ref{SpecParPlot}. Their evolution does not show a clear long-term trend, but illustrates well the striking spectral difference between the quiescent and flaring states of \src. 
In the 2011 \cxo\ observation, where the flare and quiescent states cannot be clearly defined, the spectral parameters are somehow intermediate between those of the two flux levels. The spectral parameters of all the remaining observations are instead compatible with those of the flare states.

\begin{figure*}
\centering

\resizebox{\hsize}{!}{\includegraphics[angle=0]{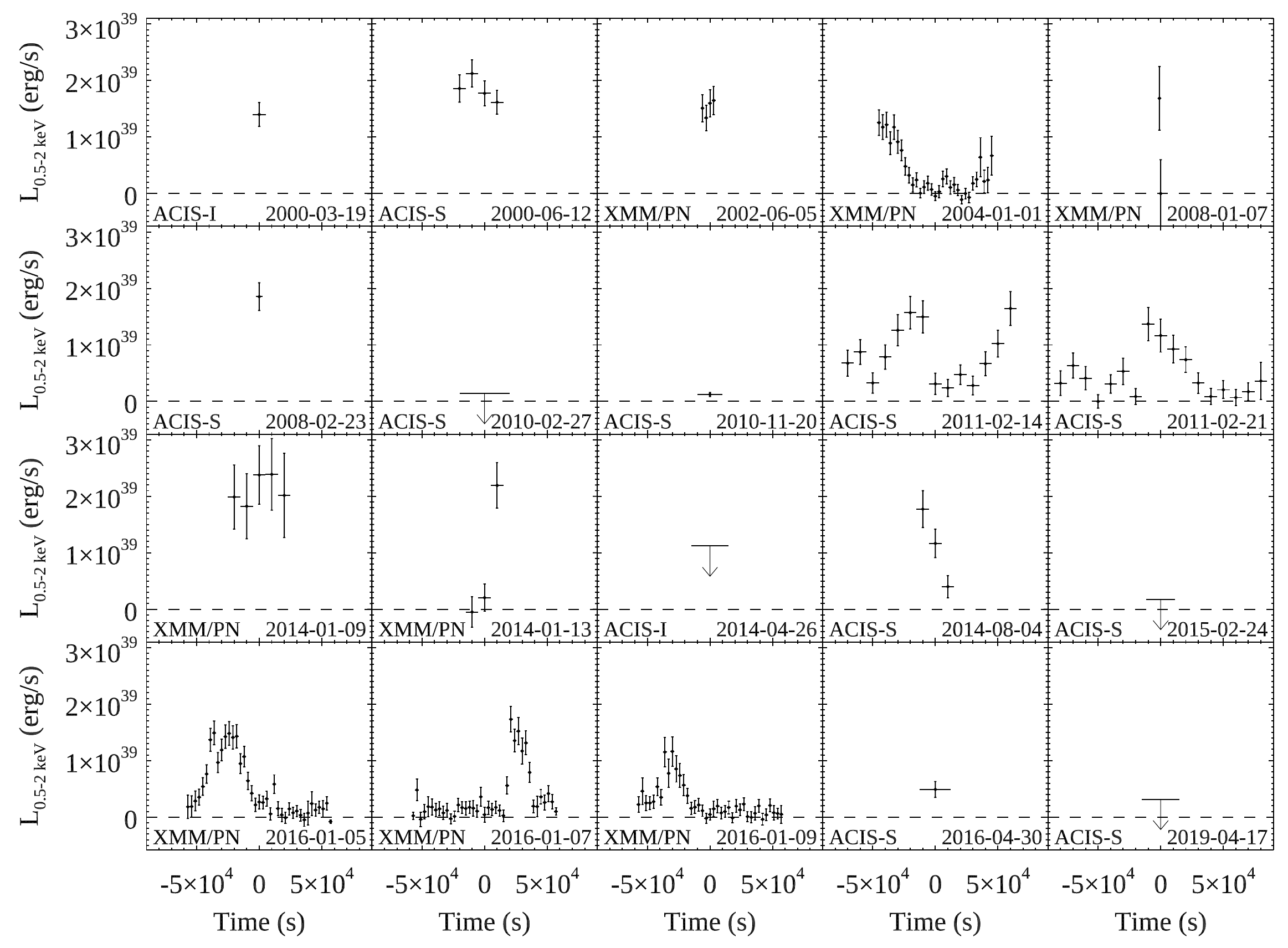}}
\caption{Light-curves and 3$\sigma$ upper limits of \src\ in luminosity units for all the \xmm\ and \cxo\ observations in Table~\ref{data}. The conversion from 0.5--2 keV count-rates into  luminosity in the 0.5--2 keV energy band was based on the best-fit parameters of the average spectrum for each observation. 
The upper limits were derived from the 0.3--2 keV count-rate limits in Table~\ref{data} assuming the best-fit spectral parameters of the first flare detected in 2016 (Table~\ref{SpecPar}). Although \citet{dage19} report a 3$\sigma$ detection in the 2019 observation, we set an upper limit because most of the counts were detected above 2 keV. The horizontal bars indicate either the bin size of the light curve (ranging from 1 to 10 ks) or the observation exposure time (for upper limits and observations with a single time bin). The start dates of the observations and the instrument are indicated in each panel.\label{alllightcurvesUL}
}
\end{figure*}

Based on the best-fit parameters of the average spectrum of each observation (including those with poorly constrained parameters, not reported in Figure~\ref{SpecParPlot}), we derived the conversion factors from the 0.5--2 keV background subtracted count rate to the observed flux in the same energy range. From these conversion factors For the \cxo\ observations where \src\ was not significantly detected, the upper limits on the source count-rates (Table~\ref{data}) were converted into 3$\sigma$ luminosity upper limits by assuming the spectral best-fit parameters of the first flare detected by \xmm\ in 2016 (Table~\ref{SpecPar}) and the ancillary response file associated to the regions from which the counts were extracted. If we instead adopted the spectral parameters of the quiescent state, depending on the detector efficiency at the lowest energies, the upper limits on luminosity would either remain virtually unchanged (e.g., in observation 11274)
or be reduced by up to $\sim$40\% (in observation 15760).

As already noted, the \ein\ and \rst\ exposures were too fragmented to display meaningful light-curves and therefore for these observations we only computed the average luminosity. For the \rst\ PSPC observation we adopted the best-fit spectral parameters shown in Figure\ref{SpecParPlot} and obtained a 0.5--2 keV luminosity of (3.0$\pm$0.3)$\times$10$^{39}$ erg\,s$^{-1}$. This is the highest flux level ever reached by \src\ (see Figure~\ref{alllightcurvesUL}).

To convert the count-rates detected during the \ein\ and \rst\ HRI observations into luminosity, we used the \texttt{WebPIMMS}\footnote{https://heasarc.gsfc.nasa.gov/cgi-bin/Tools/w3pimms/w3pimms.pl} tool, assuming a simplified spectral model: a power-law with photon index 3.7 and absorption \nh=1.55$\times$10$^{20}$ cm$^{-2}$, which, in the 0.5--2 keV energy range, satisfactorily fits the spectrum of the first flare detected by \xmm\ in 2016 ($\chi^2_{\rm red}$=1.6 for 25 d.o.f.).
The source count-rates reported in Table~\ref{data} correspond to (8.6$\pm$1.4)$\times$10$^{38}$ erg\,s$^{-1}$ for the \ein\ HRI observation and (1.1$\pm$0.2)$\times$10$^{39}$ erg\,s$^{-1}$ and (1.1$\pm$0.1)$\times$10$^{39}$ erg\,s$^{-1}$ for the \rst\ HRI observations performed in 1992 and 1994, respectively. The analysis of the latter exposure was reported also by \citet{maccarone10}, who, however, used a different spectral model, source distance and energy range to estimate the luminosity. In the same work, they report a non-detection with \ein\ in 1978, but they do not mention the detection in the HRIEXO catalog reported here. They also report the analysis of \swift/XRT observations taken in 2007, 2008 and 2010, which indicate that \src\ was not persistently in the bright state, but these exposures are too short and sparse to constrain the possible presence of flares with a duration of a few hours.

\section{Discussion}

The systematic and detailed analysis of all available \xmm\ and \cxo\ observations of \src, including also time periods with high and variable background, has allowed us to identify new episodes of source variability. In particular, since most of the variability is concentrated in the soft energy band, where the particle background contribution is less dramatic (see, e.g., Figure~\ref{PN2016lc}), we paid special attention to the choice of the most convenient energy bands for the timing analysis.

The detection of these additional X-ray flaring episodes is particularly important because they suggest a possible regular pattern (a periodicity or quasi-periodicity) and allow us to interpret more clearly the different spectral components and their evolution in the last decades.

\subsection{Implications of time variability}\label{dvar}

The analysis of the long-term variability of \src\ reported in previous works (e.g., \citealt{maccarone10,dage18}) was mostly based on the average fluxes and spectral parameters measured during each observation, without properly distinguishing flare and quiescent states. 
As a consequence, the extreme flux variability resulting from the comparison of observations with the source in different states was interpreted as a long-term trend and used as an argument against the tidal disruption scenarios \citep{maccarone10}.

Figure~\ref{alllightcurvesUL}, instead, shows that 
the long-term evolution of the source luminosity in the 0.5--2 keV energy band is apparently characterized by fast transitions between bright and faint states, with 
rather stable intensity levels of $\sim$1--2$\times$10$^{39}$ \lum\ and $\sim$2$\times$10$^{38}$ \lum, respectively. Defining as bright states all the time bins during which \src\ was detected by \xmm\ or \cxo\
above 5$\times$10$^{38}$ erg\,s$^{-1}$,
the source has been observed in the bright state for 37\% of the $\sim$1.1 Ms 
total exposure time. 

\src\ was always in a bright state during the first \cxo\ and \xmm\ observations, performed between 2000 and 2002, as well as during all the \ein\ and \rst\ exposures taken from 1979 to 1994.
Instead, \src\ was below our luminosity threshold in the two most recent \cxo\ observations. Moreover, the fraction of time spent in the bright state was 42\% between 2004 and 2015 and only 24\% during the three long \xmm\ observations performed in 2016. This suggests that before 2004 \src\ might have been persistently, or for a larger fraction of time, in the bright state.
However, the available data are too
sparse and inhomogeneous, 
with observations characterized by different exposure duration and instrumental sensitivity,
to constrain robustly a possible progressive reduction of the bright state duty cycle. 

As already noted in Section~\ref{shortVar}, the 2016 \xmm\ data are compatible with a regular flaring activity with a periodicity of $\sim$110 ks, although both the flare duration, which is a factor $\sim$2 longer for the first peak, and the peak flux, which is $\sim$30\% smaller for the last flare, display some variability. The same variability pattern is compatible also with the 2004 \xmm\ observation, which, however, starts with the source at a high flux level and has a total duration of only 90 ks. 
Therefore, in this case, we can only set a lower limit of $\sim$30 ks on the high flux state duration, which is consistent with the duration of the bright state in the first 2016 observation, and of $\sim$90 ks on the possible flare recurrence time.

Significant variability within the same luminosity limits was visible also during the two long observations performed by \cxo\ in 2011. 
If we interpret the count-rate excess detected at the end of the first observation and at the beginning of the second observation as part of a flare with a duration comparable to that of the other \src\ X-ray flares, we derive a recurrence time of $\sim$70 ks.
If we instead consider only the two flares fully contained in each exposure, 
their separation greater than one week and relatively long duration make them compatible with many possible recurrence times, including the $\sim$110 ks suggested by the 2016 \xmm\ observations (Section~\ref{shortVar}).

The other currently available observations are too short to constrain the flare duration and recurrence time. The only exception might be \xmm\ observation 0722670301 (9-10 January 2014), during which the source remained in the bright state for more time than in the longest flare detected in any other observation. However, 
this observation was affected by a high level of particle background and the telemetry buffer of the EPIC-pn quadrant hosting \src\ was saturated throughout the full observation. As a result, the live-time of this exposure is only $\sim$30\% of the total observation duration and, considering also that there are no simultaneous EPIC-MOS data, the sensitivity of this observation is lower than that of the other \xmm\ observations.

On the other hand, as described in Section~\ref{shortVar} and shown in Figure~\ref{alllightcurvesUL}, some short observations displayed rapid flux changes, which are perfectly consistent with the variability pattern described above. Moreover, we note that after 2010 \src\ has been detected about 8 times in the high flux state during $\sim$820 ks of \xmm\ and \cxo\ observations, which is compatible with the $\sim$110 ks recurrence time of the flares suggested by the most recent data. Only long, possibly uninterrupted, observations with instruments with large effective area at energies well below 1 keV, such as the EPIC-pn and, in the future, the X-IFU and WFI on-board ATHENA \citep{nandra13}, will be able to establish firmly whether the flares of \src\ occur randomly or are regularly spaced. 

\subsection{Implications of spectral analysis}\label{dspec}

The analysis of all the \src\ X-ray spectra with sufficient data quality confirms that the spectra of the short observations with high flux are compatible with those observed during the flares that were detected during the four longest \xmm\ observations (Figure~\ref{SpecParPlot}). In particular, the normalization of the disk component, which vanishes during the quiescent time periods, corresponds to an apparent inner disk radius ranging from $\sim$5,000 to $\sim$10,000 km, if an inclination of 45$^{\circ}$ is assumed. Applying equation (2) of \citet{zampieri09}, with $f=1.7$ and $b=15.6$ , these values correspond to a BH mass between $\sim$600 and $\sim$1,200 M$_{\odot}$.

Assuming that the inner disk radius $R_{\rm in}$ represents a proxy to the BH ISCO, the BH mass is related to its spin according to:

\begin{equation}
R_{\rm ISCO} = \frac{GM_{\rm BH}}{c^2}\left[ 3 + Z_2 - {\rm sign}(\xi)(3-Z_1)(3+Z_1+2Z_2)  \right]^{1/2},
\label{eq:mbh}
\end{equation}

where $Z_{1,2}$ are function of the BH dimensionless spin parameter $\xi = S_{\rm BH}/M_{\rm BH}$ \citep{1972ApJ...178..347B}. 

Assuming $R_{\rm ISCO} \equiv R_{\rm in}=6400$ km (Table~\ref{SpecPar}) and varying $S_{\rm BH} = 0-1$ leads to a range of IMBH masses of $M_{\rm IMBH} \simeq 700 - 4340$ M$_\odot$. 

We note that the spin of the IMBH is intrinsically connected with its past evolution, most likely it is inherited from the IMBH formation history. As recently discovered in a number of numerical simulations of young massive clusters, IMBHs in the mass range $100-500$ M$_\odot$ can form either from the collision and merger of massive main sequence stars \citep[e.g.][]{dicarlo21} or through the accretion of a massive star onto a stellar BH \citep[e.g.][]{2021MNRAS.501.5257R}.
In this scenario, the spin of the nascent IMBH should be either inherited from the collapse of the stellar merger remnant or from the stellar BH progenitor \citep[see e.g.][]{2019ApJ...870L..18Q}. In either cases, the IMBH spin would not deviate much from the typical values expected from stellar BHs. Note that currently there is little consensus about the BH natal spin distribution, as it strongly depends on the stellar evolution recipes adopted to model the BH formation.  

On the other hand, IMBH seeds could also grow through repeated mergers of stellar BHs, provided that the host cluster is sufficiently massive and dense. This process tends to decrease the IMBH spin, depending on the number of merging events that contributed to its build-up \citep{2020arXiv200713746A}.

A further possibility is that the IMBH growth is regulated by both processes above: first an IMBH seed with a mass $>100$ M$_\odot$ forms via stellar collisions or accretion onto stellar BHs and later the IMBH grows further via repeated BH mergers (e.g., \citealt{arcasedda21,rizzuto21}). Assuming that a population of IMBH seeds with masses in the range $100-500$ M$_\odot$ formed in globular clusters at redshift $z=2-6$ \citep{2013MNRAS.432.3250K}, recently \cite{2020arXiv200713746A} have shown that the overall population of IMBHs at redshift $z<1$ would be characterized by a mass distribution that extends between $250-1500$ M$_\odot$, with a clear peak at around $M_{\rm IMBH} \sim 750$ M$_\odot$. Therefore, the possibility to measure with high precision either the IMBH mass or its spin would provide us with crucial insights about the IMBH formation and evolution processes.  

Another possible indication that \src\ might host an IMBH is the 
high bolometric luminosity of this disk component, which ranges from 
2 to 8$\times$10$^{39}$ erg\,s$^{-1}$. 
The latter value is 
above the Eddington limit of a $\sim$10 M$_{\odot}$ BH, but corresponds to $<$10\% of the Eddington luminosity of a 1,000 M$_{\odot}$ BH.
However, this argument is not sufficient to prove the presence of an IMBH in the \gc\  GC, as demonstrated by the discovery of pulsations in ULXs (e.g., \citealt{bachetti14,israel17}),
which implies that 
they can be powered even by neutron stars.

The comparison of the spectra of the bright and faint states of \src\ unveils also a remarkable increase of the O\,\textsc{viii} emission line flux during the flares. Such a variability on hours time-scale, with no delay with respect to the disk component appearance, constrains the size of the emission region and its maximum distance from the accretion disk within 10$^{15}$ cm. The intensity of the optical [O\,\textsc{iii}] broad line is also variable from observation to observation \citep{dage19}, but HST spatially resolved spectra imply an emission region larger than a few parsecs \citep{peacock12}. Simultaneous X-ray and optical spectroscopic observations would be crucial to check whether also a fraction of the [O\,\textsc{iii}] flux might be produced in a much more compact region. 

The significantly larger flux of the power-law component in the bright state implies that also the hard X-ray emission of \src\ cannot be fully attributed to the cumulative X-ray emission of different objects in the \gc\ GC. Specifically, the same compact source responsible for the flares must be able to increase the 0.3--10 keV luminosity of its power-law spectral component by at least 5$\times$10$^{38}$ erg\,s$^{-1}$, above a similarly bright quiescent luminosity, to which other unrelated X-ray sources might contribute. 
A hard component with this luminosity can be easily produced both by a stellar and intermediate mass accreting BH.

The possibility that a variable column density might fully account for the X-ray variability of \src\
is ruled out by the simultaneous spectral fit of  the 2016 \xmm\ spectra at different intensity levels (Section~\ref{PN2016spec}). However, we showed that a combination of enhanced absorption and substantial flux reduction, possibly produced by the partial screening of the source emission by an optically thick structure, is consistent with the time-resolved spectral analysis of the same observations. Moreover,
the morphology of the 2016 light curve suggests the interpretation of the high flux states of \src\
as soft X-ray flares rather than short periods during which the X-ray emission leaks through a dense surrounding medium.

\subsection{Comparison with other systems}\label{dother}

The discovery of many episodes of bright and soft X-ray flares in the ULX \src, which might be regularly spaced in time, can be a crucial clue to interpret this unique object.
A quasi-periodic flaring behavior has been recently detected in another ULX, in the galaxy NGC~3621; in this case, the timescale is only $\sim$1 hr and the X-ray spectrum much harder \citep{motta20}. The brightest ULX known to date, HLX-1 in the ESO~243-49 galaxy, has shown a flaring variability, mainly in the soft X-ray band, with a periodicity of $\sim$1 year, but this rather regular pattern has recently changed and possibly disappeared \citep{lin20}. 

Besides, the variability of \src\ is strikingly similar to the quasi-periodic X-ray eruptions, repeating on time-scales of several hours, displayed by the active galactic nuclei (AGN) GSN~069 \citep{miniutti19} and RX~J1301.9+2747 \citep{giustini20}, as well as by two nuclear X-ray sources in previously quiescent galaxies \citep{arcodia21} and a tidal disruption event (TDE;
\citealt{chakraborty21}).
Their origin might be related to disk instabilities (e.g., \citealt{sniegowska20}), the partial disruption of an orbiting star   \citep{king20,sheng21}, the self-lensing of a supermassive black hole binary system \citep{ingram21}, or  transient accretion in extreme-mass ratio inspirals \citep{metzger21,zhao21}. Although their luminosity is much larger, we note that also these flaring 
X-ray sources are characterized by a very soft spectrum. Moreover, the regular variability of GSN~069 started after several years of bright persistent emission, as it might have happened in the case of \src.

A Galactic X-ray source with interesting similarities with \src, is the BH candidate X9 at the center of the GC 47~Tuc \citep{mj15}: it displays flaring X-ray variability on an apparently regular timescale of $\sim$1 week and its X-ray spectrum is characterized by oxygen emission lines \citep{bahramian17}. Its luminosity, however, ranges from 10$^{33}$ to 10$^{34}$ \lum, which is more than 5 orders of magnitude below that of \src, which, in turn, is more than 3 orders of magnitude less luminous than the QPEs observed in the galactic nuclei \citep{miniutti19,giustini20, arcodia21,chakraborty21}.

 A particularly intriguing possibility is that these accreting sources, all located in extremely crowded regions (at the very center of GCs or galaxies), might share a common mechanism responsible for the modulation of their X-ray luminosity on different time-scales, perhaps the interaction with a star, very likely on a highly eccentric orbit, that might perturb a pre-existing disk  or provide matter through partial tidal disruption.
 In addition, the partial tidal disruption of a star orbiting at short distance from the supermassive BH at the center of our Galaxy has been 
proposed as the possible cause of the X-ray and NIR flaring activity of Sgr\,A$^*$ \citep{leibowitz20}.
 
Finally, \citet{maccarone05} has proposed to interpret a small class of X-ray transients in the NGC 4697 galaxy \citep{sivakoff05} as periodic accretion episodes of eccentric binary systems in GCs. More recently, after the discovery of other extra-galactic X-ray transients with similar properties \citep{jonker13,Irwin16}, \citet{shen19} has interpreted these transients as partial tidal disruptions of WDs in eccentric (or parabolic) orbits around IMBHs. 
In the next section, we explore a similar scenario also for \src.
Although the spectra of these transients are harder, their luminosities are larger, and their light-curves have asymmetric profiles, we note that also
TDEs display a rather rich X-ray phenomenology, including both soft and hard spectra (see \citealt{saxton20} for a review).
Due to its intermediate variability time-scale and luminosity 
among all the phenomena reported above, \src\ might be a crucial object to clarify this unifying scenario.

\subsection{Partial tidal disruption scenario}\label{dTDE}

We 
explore the possibility that the recurrent X-ray flares from XMMU J122939.7+075333
might be produced by an extreme mass transfer system, where the orbit of the secondary star is highly eccentric, so that the star fills its Roche lobe only at pericenter, losing a small amount of mass at each passage.
Due to the high eccentricity implied (see below), we use the formalism developed for partial tidal disruption events (pTDE), where a star on a parabolic orbit is disrupted by a massive black hole (see, e.g., \citealt{rees88,guillochon13}). 

We define the stellar mass $M_*=m_*\text{M}_\odot$ and radius $R_*=r_*\text{R}_{\odot}$, and the BH mass $M_{\rm BH}$. The Roche lobe is given by \citep{eggleton83}
\begin{equation}
    R_{\rm L}\sim 0.46\times \left( \frac{M_*}{M_{\rm BH}}\right)^{1/3}a(1-e),
    \label{eq:RL}
\end{equation}
where $a$ is the semi-major axis, $e$ is the orbital eccentricity, and the pericenter is $r_{\rm p}=a(1-e)$. 
The strength of this kind of encounters is usually quantified through a dimensionless parameter called \emph{penetration factor}, defined as\footnote{This is an approximated equality since there is also a factor of a few depending on some specific feature of the TDE, like the internal structure of the star involved in the disruption.} 
\begin{equation}
\beta=r_{\rm t}/r_{\rm p},
\end{equation}
where $r_{\rm t}\simeq(M_{\rm BH}/M_*)^{1/3}R_*$ is the maximum distance at which the star is fully disrupted by the BH (a.k.a. tidal radius). Generally speaking, if $\beta \lesssim 1$ just a fraction of the stellar mass is stripped away and accreted onto the BH, and thus we have a pTDE. 
In terms of the Roche lobe, this means that at the pericenter
\begin{equation}
    R_*\simeq R_{\rm L}\beta,
    \label{eq:lobe}
\end{equation}
hence if $\beta \lesssim 1$ the lobe is just partially filled by the stellar material.
\citet{guillochon13} show that, when $\beta$ drops below $\sim 0.5$, the mass transfer 
virtually stops.

We estimate the value of $\beta$ from the total mass accreted during each flare using the correlation obtained from hydrodynamical simulations of pTDE \citep{guillochon13}. 
The total energy emitted by this source during each of the three 2016 XMM observations is 
  \begin{equation}
      E_{1}\sim 1.2\times 10^{44}\,\text{erg}, E_{2}\sim 6.8\times 10^{43}\,\text{erg}, E_{3}\sim 2.5\times 10^{43}\,\text{erg},
  \end{equation} 
respectively. Assuming an accretion efficiency $\eta=0.1$, we obtain that the total mass accreted onto the BH during each flare in units of solar masses is of the order
  \begin{equation}
  \Delta M_{1}\sim 6.7\times 10^{-10}, \Delta M_{2}\sim 3.8\times 10^{-10}, \Delta M_{3} \sim 1.4\times 10^{-10}.
  \end{equation}
The large abundance of oxygen observed suggests that the star involved in the event is a WD. Considering a WD with mass in the range $0.1M_\odot \leq M_{*}\leq 1.4M_\odot$, we have that the ratio of the accreted mass over the stellar mass is in the following interval
  \begin{equation}
      10^{-10}\lesssim \frac{\Delta M}{M_*}\lesssim 10^{-9} .
  \end{equation}
Such small values of the ratio imply that the BH-WD encounters must be very grazing and so that the $\beta$ parameter needs to be very close to $\sim 0.5$, with a small dependence on the stellar structure, modeled as a polytropic sphere with index $\gamma$ (\citealt{guillochon13}, equations A3-A9-A10 Appendix A). In particular, we obtain $\beta\approx 0.45$ for $\gamma=4/3$ and $\beta\approx 0.5$ for $\gamma=5/3$.

 Note that, since the accreted mass drops steeply when $\beta$ becomes smaller than $\simeq 0.5$, the resulting  $\beta$ is very insensitive to the actual value of accreted mass. In other words, since $\Delta M/M_*$ is very small, $\beta$ has to be very close to $\simeq 0.45$ (or 0.5, depending on $\gamma$), with differences occurring only at the $0.1$\% level. 

 \begin{figure}
     \centering
     \includegraphics[scale=0.45]{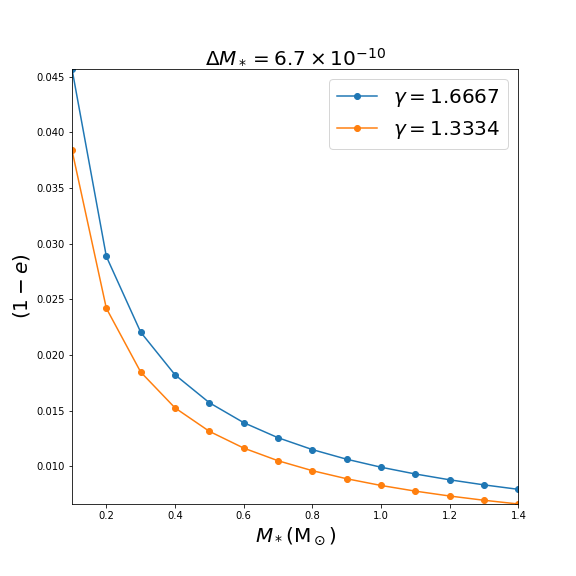}
     \caption{$(1-e)$ vs $M_*$ for the first 2016 XMM observation.}
     \label{fig:fig1}
 \end{figure}

 To estimate also the other orbital parameters of the event, we can proceed in the following way. From the best fit of the inner disk radius $R_{\rm in}\sim 6400\,\text{km}$ (Table~\ref{SpecPar}) and equation~\ref{eq:mbh}, assuming 
 a Schwarzschild BH, we obtain 
    $M_{\rm BH}\approx 700\,\text{M}_{\odot}$.
 Using this information together with the estimated period of $112\,\text{ks}$ suggested by the 2016 observations, we obtain the semi-major axis of the stellar orbit
 \begin{equation}
     a\approx 3\times 10^{12}\,\text{cm}.
     \label{eq:smaxis}
 \end{equation}

Finally, considering for a WD the following mass-radius relation 
 \begin{equation}
     \frac{R_*}{R_\odot}\approx 10^{-2}\left(\frac{M_{*}}{M_\odot}\right)^{-1/3},
     \label{eq:radius}
 \end{equation}
 we can re-write equation \eqref{eq:lobe} as
 \begin{equation}
    (1-e) \approx 0.0045 m_*^{-2/3}\beta^{-1}.
 \end{equation}
We note that this relation holds also if we consider a stellar BH instead of an IMBH, because from equations \ref{eq:RL} and \ref{eq:lobe} we see that
\begin{equation}
    (1-e)\sim \frac{1}{a}\times \left(\frac{M_{\rm BH}}{M_*}\right)^{1/3}\frac{R_*}{0.46\beta}.
\end{equation}
Hence, this relation depends on the ratio $M_{\rm BH}^{1/3}/a$, that is constant since we take the period as a fixed value derived from observations.
We thus conclude that, independently of the stellar mass, the orbit must be highly eccentric, as shown in Figure \ref{fig:fig1}. During the three observations, the variation of the orbital eccentricity, for each stellar mass inside the range specified previously, is of order $10^{-4}$.

According to our study, the three quasi-periodic flares seen by \xmm\ in 2016 are compatible with a pTDE, if the following conditions are fulfilled:
i) very grazing encounters, with $0.45 \lesssim \beta \lesssim 0.55$ depending on the internal structure of the star and ii) highly eccentric orbits, $e>0.95$.

We expect the star involved in this event to be a WD. Indeed, the presence of a WD is supported by observations, since abundant oxygen has been revealed. Furthermore, the small fraction of accreted mass per orbit ($\sim 10^{-10}$) is in agreement with the hypothesis of a stellar object characterized by large compactness. 

The tidal disruption of a WD by an IMBH was considered by \citet{clausen11} to explain the broad [O\,\textsc{iii}] line observed in the optical spectra of \src: by modeling the photoionization of the unbound debris by the accretion flare, its luminosity and width could be very well reproduced, but this possibility was discarded due to the too short timescale expected in case of a full stellar disruption. This limitation, however, is not relevant for the repeating partial disruption that we propose to explain the phenomenology displayed by \src.

Our results are consistent with the work presented in \citet{zalamea10}, who investigated the gravitational emission from tidally stripped WDs.
They divide the process into two phases: the first one is longer and characterized by (slow) gravitational wave in-spiraling, while the second one is shorter and terminates with the detonation of the WD. During the first stage, the periodic mass loss occurring at the pericenter is very small and thus the overall mass of the WD is roughly unchanged. This phase, which is characterized by periodic electromagnetic signals from the system, continues for a number of orbits given by \citep{zalamea10,shen19} 
\begin{align}
\mathcal{N}_{\rm orb}\approx 10^{3}\times \left(\frac{M_{\rm BH}}{10^5\text{M}_{\odot}}\right)^{-2/5}.
\end{align}
If we apply this estimate to our case, we have $\mathcal{N}_{\rm orb}\approx 7000$ orbits, corresponding to $\approx$25 years for an orbital period of 112 ks. This timescale is compatible with the time interval over which we observed the recurrent X-ray flares.

Unlike \citet{zalamea10}, \citet{McLeod2014} simulate a pTDE of a WD by a massive BH ($\sim 10^{5}\,\text{M}_{\odot}$) and they find that this process lasts 
a few tens of orbits before the full disruption of the star,
while for \citet{zalamea10} the process should be much longer. The only way to alleviate this tension is performing further simulations of this scenario.

We note that the relationship between the amount of accreted mass and the penetration factor that we use \citep{guillochon13} is based on numerical simulations of pTDE of main sequence stars on parabolic orbits.\footnote{Recently, \citet{Wang21} have also simulated pTDEs, but assuming stellar BHs, lighter than the one we consider for this work.}  Although it is 
the easiest assumption to extrapolate these numbers for the case at hand, a proper set of numerical simulation will be performed in future to estimate more precisely how the orbital parameters change for the case of interest.

Finally, a curious feature of XMMU J122939.7+075333 is that 
before 2004
it was always observed 
in a high luminosity state. 
However, after a while, it started to alternate between periods of high luminosity and periods of quiescence.
This behavior could be explained assuming that initially the outer layers of the WD were stripped away and for some time they were continuously accreted on the BH. The surviving core might then have returned to the pericenter, giving rise to the recurrent flares that we now observe. Another intriguing possibility is that the WD orbit has become wider and/or more eccentric with time, switching from a persistently accreting X-ray binary to a system where the donor star loses mass only close to the periastron. Although this kind of evolution is predicted by evolution models of eccentric binaries where the accretor is more massive than the donor (see, e.g., \citealt{dosopoulou16}), in order to obtain a large period increase in only a few tens of years, a more complex scenario might be envisaged (e.g., a third object might drive the orbital evolution through the Kozai--Lidov mechanism \citealt{Kozai62,lidov62}).
This hypothesis can be tested by future observations, which might confirm the flare recurrence and its possible increase in frequency with time.

\section{Summary and conclusions}

The systematic analysis of the X-ray data currently available on the ULX \src\ has allowed us to identify for the first time a possibly regular flaring behavior. The spectral analysis of these X-ray flares suggests that they can be produced by accretion onto a $\sim$10$^3$ M$_\odot$ BH and that the donor star might be oxygen rich. The properties of the peculiar optical counterpart support the latter conclusion and indicate a WD as the likely origin of the accreting matter. We therefore propose that the recurrent flares might be produced by the partial disruption of a WD in a very eccentric orbit around an IMBH. Being located in a GC, the extreme eccentricity required by our model can originate from the close encounter between the two objects, which might have occurred only a few decades ago, as possibly indicated by the higher, possibly persistent, luminosity registered in the oldest observations and the gradual reduction of the flare duty cycle in the most recent ones.

This scenario resembles the ones proposed by many authors \citep{king20, sheng21, arcodia21, chakraborty21} to interpret the QPEs recently detected in the nuclear regions of galaxies, very likely hosting supermassive BHs with relatively small masses ($\sim$10$^5$--10$^6$ M$_\odot$). If the regular flaring pattern of \src\ were confirmed by future observations, it would be a scaled-down example of the QPE phenomenon, 
in a different environment, characterized, however, 
by a similarly large stellar density. We finally note that this peculiar accretion regime might have already been observed in other GC X-ray sources, such as the X9 source in the center of the 47 Tuc Galactic GC \citep{mj15} and several extra-galactic X-ray transients \citep{sivakoff05,jonker13,Irwin16}.

\begin{acknowledgements}
The scientific results reported in this article are based on 
observations obtained with \xmm, an ESA science mission with 
instruments and contributions directly funded by ESA Member States and NASA, and on data obtained from the \cxo\ Data Archive and has made use of the software package CIAO provided by the \cxo\ X-ray Center (CXC). We acknowledge the computing centers of INAF (Osservatorio Astronomico di Trieste / Osservatorio Astrofisico di Catania), under the coordination of the CHIPP project, for the availability of computing resources and support. We acknowledge financial support from ASI under ASI/INAF agreement N.2017-14.H.0 AT and PE acknowledge financial support from the Italian Ministry for University and Research, through grant 2017LJ39LM (UNIAM). MAS acknowledges the Alexander von Humboldt Foundation for the financial support provided in the framework of the research program "The evolution of black holes from stellar to galactic scales", the Volkswagen Foundation Trilateral Partnership project No. I/97778 “Dynamical Mechanisms of Accretion
in Galactic Nuclei”, and the Sonderforschungsbereich SFB 881 "The Milky Way System" – Project-ID 138713538 – funded by the German Research Foundation (DFG).

\end{acknowledgements}
\bibliographystyle{aa} 
\bibliography{biblio} 

\begin{thebibliography}{67}
\expandafter\ifx\csname natexlab\endcsname\relax\def\natexlab#1{#1}\fi

\bibitem[{{Arca Sedda} {et~al.}(2021){Arca Sedda}, {Amaro Seoane}, \&
  {Chen}}]{2020arXiv200713746A}
{Arca Sedda}, M., {Amaro Seoane}, P., \& {Chen}, X. 2021, \aap, 652, A54

\bibitem[{{Arca Sedda} {et~al.}(2018){Arca Sedda}, {Askar}, \&
  {Giersz}}]{arcasedda18}
{Arca Sedda}, M., {Askar}, A., \& {Giersz}, M. 2018, \mnras, 479, 4652

\bibitem[{{Arca-Sedda} {et~al.}(2021){Arca-Sedda}, {Rizzuto}, {Naab},
  {Ostriker}, {Giersz}, \& {Spurzem}}]{arcasedda21}
{Arca-Sedda}, M., {Rizzuto}, F.~P., {Naab}, T., {et~al.} 2021, \apj, 920, 128

\bibitem[{{Arcodia} {et~al.}(2021){Arcodia}, {Merloni}, {Nandra}, {Buchner},
  {Salvato}, {Pasham}, {Remillard}, {Comparat}, {Lamer}, {Ponti}, {Malyali},
  {Wolf}, {Arzoumanian}, {Bogensberger}, {Buckley}, {Gendreau}, {Gromadzki},
  {Kara}, {Krumpe}, {Markwardt}, {Ramos-Ceja}, {Rau}, {Schramm}, \&
  {Schwope}}]{arcodia21}
{Arcodia}, R., {Merloni}, A., {Nandra}, K., {et~al.} 2021, \nat, 592, 704

\bibitem[{{Bachetti} {et~al.}(2014){Bachetti}, {Harrison}, {Walton},
  {Grefenstette}, {Chakrabarty}, {F{\"u}rst}, {Barret}, {Beloborodov}, {Boggs},
  {Christensen}, {Craig}, {Fabian}, {Hailey}, {Hornschemeier}, {Kaspi},
  {Kulkarni}, {Maccarone}, {Miller}, {Rana}, {Stern}, {Tendulkar}, {Tomsick},
  {Webb}, \& {Zhang}}]{bachetti14}
{Bachetti}, M., {Harrison}, F.~A., {Walton}, D.~J., {et~al.} 2014, \nat, 514,
  202

\bibitem[{{Bahramian} {et~al.}(2017){Bahramian}, {Heinke}, {Tudor},
  {Miller-Jones}, {Bogdanov}, {Maccarone}, {Knigge}, {Sivakoff}, {Chomiuk},
  {Strader}, {Garcia}, \& {Kallman}}]{bahramian17}
{Bahramian}, A., {Heinke}, C.~O., {Tudor}, V., {et~al.} 2017, \mnras, 467, 2199

\bibitem[{{Bardeen} {et~al.}(1972){Bardeen}, {Press}, \&
  {Teukolsky}}]{1972ApJ...178..347B}
{Bardeen}, J.~M., {Press}, W.~H., \& {Teukolsky}, S.~A. 1972, \apj, 178, 347

\bibitem[{{Chakraborty} {et~al.}(2021){Chakraborty}, {Kara}, {Masterson},
  {Giustini}, {Miniutti}, \& {Saxton}}]{chakraborty21}
{Chakraborty}, J., {Kara}, E., {Masterson}, M., {et~al.} 2021, \apjl, 921, L40

\bibitem[{{Clausen} \& {Eracleous}(2011)}]{clausen11}
{Clausen}, D. \& {Eracleous}, M. 2011, \apj, 726, 34

\bibitem[{{Dage} {et~al.}(2018){Dage}, {Zepf}, {Bahramian}, {Kundu},
  {Maccarone}, \& {Peacock}}]{dage18}
{Dage}, K.~C., {Zepf}, S.~E., {Bahramian}, A., {et~al.} 2018, \apj, 862, 108

\bibitem[{{Dage} {et~al.}(2019){Dage}, {Zepf}, {Bahramian}, {Strader},
  {Maccarone}, {Peacock}, {Kundu}, {Steele}, \& {Britt}}]{dage19}
{Dage}, K.~C., {Zepf}, S.~E., {Bahramian}, A., {et~al.} 2019, \mnras, 489, 4783

\bibitem[{{De Luca} {et~al.}(2021){De Luca}, {Salvaterra}, {Belfiore},
  {Carpano}, {D'Agostino}, {Haberl}, {Israel}, {Law-Green}, {Lisini},
  {Marelli}, {Novara}, {Read}, {Rodriguez-Castillo}, {Rosen}, {Salvetti},
  {Tiengo}, {Vianello}, {Watson}, {Delvaux}, {Dickens}, {Esposito}, {Greiner},
  {H{\"a}mmerle}, {Kreikenbohm}, {Kreykenbohm}, {Oertel}, {Pizzocaro}, {Pye},
  {Sandrelli}, {Stelzer}, {Wilms}, \& {Zagaria}}]{deluca21extras}
{De Luca}, A., {Salvaterra}, R., {Belfiore}, A., {et~al.} 2021, \aap, 650, A167

\bibitem[{{Di Carlo} {et~al.}(2021){Di Carlo}, {Mapelli}, {Pasquato},
  {Rastello}, {Ballone}, {Dall'Amico}, {Giacobbo}, {Iorio}, {Spera},
  {Torniamenti}, \& {Haardt}}]{dicarlo21}
{Di Carlo}, U.~N., {Mapelli}, M., {Pasquato}, M., {et~al.} 2021, \mnras, 507,
  5132

\bibitem[{{Dosopoulou} \& {Kalogera}(2016)}]{dosopoulou16}
{Dosopoulou}, F. \& {Kalogera}, V. 2016, \apj, 825, 71

\bibitem[{{Eggleton}(1983)}]{eggleton83}
{Eggleton}, P.~P. 1983, \apj, 268, 368

\bibitem[{{Garmire} {et~al.}(2003){Garmire}, {Bautz}, {Ford}, {Nousek}, \&
  {Ricker}}]{garmire03}
{Garmire}, G.~P., {Bautz}, M.~W., {Ford}, P.~G., {Nousek}, J.~A., \& {Ricker},
  Jr., G.~R. 2003, in Proceedings of the SPIE., Vol. 4851, X-Ray and Gamma-Ray
  Telescopes and Instruments for Astronomy., ed. J.~E. {Truemper} \& H.~D.
  {Tananbaum} (SPIE, Bellingham), 28--44

\bibitem[{{Giacconi} {et~al.}(1979){Giacconi}, {Branduardi}, {Briel},
  {Epstein}, {Fabricant}, {Feigelson}, {Forman}, {Gorenstein}, {Grindlay},
  {Gursky}, {Harnden}, {Henry}, {Jones}, {Kellogg}, {Koch}, {Murray},
  {Schreier}, {Seward}, {Tananbaum}, {Topka}, {Van Speybroeck}, {Holt},
  {Becker}, {Boldt}, {Serlemitsos}, {Clark}, {Canizares}, {Markert}, {Novick},
  {Helfand}, \& {Long}}]{giacconi79}
{Giacconi}, R., {Branduardi}, G., {Briel}, U., {et~al.} 1979, \apj, 230, 540

\bibitem[{{Giesers} {et~al.}(2019){Giesers}, {Kamann}, {Dreizler}, {Husser},
  {Askar}, {G{\"o}ttgens}, {Brinchmann}, {Latour}, {Weilbacher}, {Wendt}, \&
  {Roth}}]{giesers19}
{Giesers}, B., {Kamann}, S., {Dreizler}, S., {et~al.} 2019, \aap, 632, A3

\bibitem[{{Giustini} {et~al.}(2020){Giustini}, {Miniutti}, \&
  {Saxton}}]{giustini20}
{Giustini}, M., {Miniutti}, G., \& {Saxton}, R.~D. 2020, \aap, 636, L2

\bibitem[{{Guillochon} \& {Ramirez-Ruiz}(2013)}]{guillochon13}
{Guillochon}, J. \& {Ramirez-Ruiz}, E. 2013, \apj, 767, 25

\bibitem[{{Ingram} {et~al.}(2021){Ingram}, {Motta}, {Aigrain}, \&
  {Karastergiou}}]{ingram21}
{Ingram}, A., {Motta}, S.~E., {Aigrain}, S., \& {Karastergiou}, A. 2021,
  \mnras, 503, 1703

\bibitem[{{Irwin} {et~al.}(2010){Irwin}, {Brink}, {Bregman}, \&
  {Roberts}}]{irwin10}
{Irwin}, J.~A., {Brink}, T.~G., {Bregman}, J.~N., \& {Roberts}, T.~P. 2010,
  \apjl, 712, L1

\bibitem[{{Irwin} {et~al.}(2016){Irwin}, {Maksym}, {Sivakoff}, {Romanowsky},
  {Lin}, {Speegle}, {Prado}, {Mildebrath}, {Strader}, {Liu}, \&
  {Miller}}]{Irwin16}
{Irwin}, J.~A., {Maksym}, W.~P., {Sivakoff}, G.~R., {et~al.} 2016, \nat, 538,
  356

\bibitem[{{Israel} {et~al.}(2017){Israel}, {Belfiore}, {Stella}, {Esposito},
  {Casella}, {De Luca}, {Marelli}, {Papitto}, {Perri}, {Puccetti}, {Castillo},
  {Salvetti}, {Tiengo}, {Zampieri}, {D'Agostino}, {Greiner}, {Haberl},
  {Novara}, {Salvaterra}, {Turolla}, {Watson}, {Wilms}, \& {Wolter}}]{israel17}
{Israel}, G.~L., {Belfiore}, A., {Stella}, L., {et~al.} 2017, Science, 355, 817

\bibitem[{{Jonker} {et~al.}(2013){Jonker}, {Glennie}, {Heida}, {Maccarone},
  {Hodgkin}, {Nelemans}, {Miller-Jones}, {Torres}, \& {Fender}}]{jonker13}
{Jonker}, P.~G., {Glennie}, A., {Heida}, M., {et~al.} 2013, \apj, 779, 14

\bibitem[{{Joseph} {et~al.}(2015){Joseph}, {Maccarone}, {Kraft}, \&
  {Sivakoff}}]{joseph15}
{Joseph}, T.~D., {Maccarone}, T.~J., {Kraft}, R.~P., \& {Sivakoff}, G.~R. 2015,
  \mnras, 447, 1460

\bibitem[{{Katz} \& {Ricotti}(2013)}]{2013MNRAS.432.3250K}
{Katz}, H. \& {Ricotti}, M. 2013, \mnras, 432, 3250

\bibitem[{{King}(2020)}]{king20}
{King}, A. 2020, \mnras, 493, L120

\bibitem[{{Kozai}(1962)}]{Kozai62}
{Kozai}, Y. 1962, \aj, 67, 591

\bibitem[{{Leibowitz}(2020)}]{leibowitz20}
{Leibowitz}, E. 2020, \apj, 896, 74

\bibitem[{{Lidov}(1962)}]{lidov62}
{Lidov}, M.~L. 1962, \planss, 9, 719

\bibitem[{{Lin} {et~al.}(2020){Lin}, {Hu}, {Li}, {Takata}, {Yen}, {Kwak},
  {Kim}, \& {Kong}}]{lin20}
{Lin}, L. C.-C., {Hu}, C.-P., {Li}, K.-L., {et~al.} 2020, \mnras, 491, 5682

\bibitem[{{Maccarone}(2005)}]{maccarone05}
{Maccarone}, T.~J. 2005, \mnras, 364, 971

\bibitem[{{Maccarone} {et~al.}(2007){Maccarone}, {Kundu}, {Zepf}, \&
  {Rhode}}]{maccarone07}
{Maccarone}, T.~J., {Kundu}, A., {Zepf}, S.~E., \& {Rhode}, K.~L. 2007, \nat,
  445, 183

\bibitem[{{Maccarone} {et~al.}(2010){Maccarone}, {Kundu}, {Zepf}, \&
  {Rhode}}]{maccarone10}
{Maccarone}, T.~J., {Kundu}, A., {Zepf}, S.~E., \& {Rhode}, K.~L. 2010, \mnras,
  409, L84

\bibitem[{{MacLeod} {et~al.}(2014){MacLeod}, {Goldstein}, {Ramirez-Ruiz},
  {Guillochon}, \& {Samsing}}]{McLeod2014}
{MacLeod}, M., {Goldstein}, J., {Ramirez-Ruiz}, E., {Guillochon}, J., \&
  {Samsing}, J. 2014, \apj, 794, 9

\bibitem[{{Metzger} {et~al.}(2021){Metzger}, {Stone}, \& {Gilbaum}}]{metzger21}
{Metzger}, B.~D., {Stone}, N.~C., \& {Gilbaum}, S. 2021, arXiv e-prints,
  arXiv:2107.13015

\bibitem[{{Miller-Jones} {et~al.}(2015){Miller-Jones}, {Strader}, {Heinke},
  {Maccarone}, {van den Berg}, {Knigge}, {Chomiuk}, {Noyola}, {Russell},
  {Seth}, \& {Sivakoff}}]{mj15}
{Miller-Jones}, J.~C.~A., {Strader}, J., {Heinke}, C.~O., {et~al.} 2015,
  \mnras, 453, 3918

\bibitem[{{Miniutti} {et~al.}(2019){Miniutti}, {Saxton}, {Giustini}, {Alexand
  er}, {Fender}, {Heywood}, {Monageng}, {Coriat}, {Tzioumis}, {Read}, {Knigge},
  {Gandhi}, {Pretorius}, \& {Ag{\'\i}s-Gonz{\'a}lez}}]{miniutti19}
{Miniutti}, G., {Saxton}, R.~D., {Giustini}, M., {et~al.} 2019, \nat, 573, 381

\bibitem[{{Morscher} {et~al.}(2015){Morscher}, {Pattabiraman}, {Rodriguez},
  {Rasio}, \& {Umbreit}}]{morscher15}
{Morscher}, M., {Pattabiraman}, B., {Rodriguez}, C., {Rasio}, F.~A., \&
  {Umbreit}, S. 2015, \apj, 800, 9

\bibitem[{{Motta} {et~al.}(2020){Motta}, {Marelli}, {Pintore}, {Esposito},
  {Salvaterra}, {De Luca}, {Israel}, {Tiengo}, \& {Castillo}}]{motta20}
{Motta}, S.~E., {Marelli}, M., {Pintore}, F., {et~al.} 2020, \apj, 898, 174

\bibitem[{{Nandra} {et~al.}(2013){Nandra}, {Barret}, {Barcons}, {Fabian}, {den
  Herder}, {Piro}, {Watson}, {Adami}, {Aird}, {Afonso}, {Alexander},
  {Argiroffi}, {Amati}, {Arnaud}, {Atteia}, {Audard}, {Badenes}, {Ballet},
  {Ballo}, {Bamba}, {Bhardwaj}, {Stefano Battistelli}, {Becker}, {De Becker},
  {Behar}, {Bianchi}, {Biffi}, {B{\^\i}rzan}, {Bocchino}, {Bogdanov}, {Boirin},
  {Boller}, {Borgani}, {Borm}, {Bouch{\'e}}, {Bourdin}, {Bower}, {Braito},
  {Branchini}, {Branduardi-Raymont}, {Bregman}, {Brenneman}, {Brightman},
  {Br{\"u}ggen}, {Buchner}, {Bulbul}, {Brusa}, {Bursa}, {Caccianiga},
  {Cackett}, {Campana}, {Cappelluti}, {Cappi}, {Carrera}, {Ceballos},
  {Christensen}, {Chu}, {Churazov}, {Clerc}, {Corbel}, {Corral}, {Comastri},
  {Costantini}, {Croston}, {Dadina}, {D'Ai}, {Decourchelle}, {Della Ceca},
  {Dennerl}, {Dolag}, {Done}, {Dovciak}, {Drake}, {Eckert}, {Edge}, {Ettori},
  {Ezoe}, {Feigelson}, {Fender}, {Feruglio}, {Finoguenov}, {Fiore}, {Galeazzi},
  {Gallagher}, {Gandhi}, {Gaspari}, {Gastaldello}, {Georgakakis},
  {Georgantopoulos}, {Gilfanov}, {Gitti}, {Gladstone}, {Goosmann}, {Gosset},
  {Grosso}, {Guedel}, {Guerrero}, {Haberl}, {Hardcastle}, {Heinz}, {Alonso
  Herrero}, {Herv{\'e}}, {Holmstrom}, {Iwasawa}, {Jonker}, {Kaastra}, {Kara},
  {Karas}, {Kastner}, {King}, {Kosenko}, {Koutroumpa}, {Kraft}, {Kreykenbohm},
  {Lallement}, {Lanzuisi}, {Lee}, {Lemoine-Goumard}, {Lobban}, {Lodato},
  {Lovisari}, {Lotti}, {McCharthy}, {McNamara}, {Maggio}, {Maiolino}, {De
  Marco}, {de Martino}, {Mateos}, {Matt}, {Maughan}, {Mazzotta}, {Mendez},
  {Merloni}, {Micela}, {Miceli}, {Mignani}, {Miller}, {Miniutti}, {Molendi},
  {Montez}, {Moretti}, {Motch}, {Naz{\'e}}, {Nevalainen}, {Nicastro}, {Nulsen},
  {Ohashi}, {O'Brien}, {Osborne}, {Oskinova}, {Pacaud}, {Paerels}, {Page},
  {Papadakis}, {Pareschi}, {Petre}, {Petrucci}, {Piconcelli}, {Pillitteri},
  {Pinto}, {de Plaa}, {Pointecouteau}, {Ponman}, {Ponti}, {Porquet}, {Pounds},
  {Pratt}, {Predehl}, {Proga}, {Psaltis}, {Rafferty}, {Ramos-Ceja}, {Ranalli},
  {Rasia}, {Rau}, {Rauw}, {Rea}, {Read}, {Reeves}, {Reiprich}, {Renaud},
  {Reynolds}, {Risaliti}, {Rodriguez}, {Rodriguez Hidalgo}, {Roncarelli},
  {Rosario}, {Rossetti}, {Rozanska}, {Rovilos}, {Salvaterra}, {Salvato}, {Di
  Salvo}, {Sanders}, {Sanz-Forcada}, {Schawinski}, {Schaye}, {Schwope},
  {Sciortino}, {Severgnini}, {Shankar}, {Sijacki}, {Sim}, {Schmid}, {Smith},
  {Steiner}, {Stelzer}, {Stewart}, {Strohmayer}, {Str{\"u}der}, {Sun}, {Takei},
  {Tatischeff}, {Tiengo}, {Tombesi}, {Trinchieri}, {Tsuru}, {Ud-Doula},
  {Ursino}, {Valencic}, {Vanzella}, {Vaughan}, {Vignali}, {Vink}, {Vito},
  {Volonteri}, {Wang}, {Webb}, {Willingale}, {Wilms}, {Wise}, {Worrall},
  {Young}, {Zampieri}, {In't Zand}, {Zane}, {Zezas}, {Zhang}, \&
  {Zhuravleva}}]{nandra13}
{Nandra}, K., {Barret}, D., {Barcons}, X., {et~al.} 2013, arXiv e-prints,
  arXiv:1306.2307

\bibitem[{{Peacock} {et~al.}(2012){Peacock}, {Zepf}, {Kundu}, {Maccarone},
  {Rhode}, {Salzer}, {Waters}, {Ciardullo}, {Gronwall}, \& {Stern}}]{peacock12}
{Peacock}, M.~B., {Zepf}, S.~E., {Kundu}, A., {et~al.} 2012, \apj, 759, 126

\bibitem[{{Pfeffermann} {et~al.}(1987){Pfeffermann}, {Briel}, {Hippmann},
  {Kettenring}, {Metzner}, {Predehl}, {Reger}, {Stephan}, {Zombeck},
  {Chappell}, \& {Murray}}]{pfeffermann87}
{Pfeffermann}, E., {Briel}, U.~G., {Hippmann}, H., {et~al.} 1987, in SPIE
  Conference Series, Vol. 733, Soft X-ray optics and technology. Edited by
  E.-E.~Koch \& G.~Schmahl (SPIE, Bellingham), 519--532

\bibitem[{{Qin} {et~al.}(2019){Qin}, {Marchant}, {Fragos}, {Meynet}, \&
  {Kalogera}}]{2019ApJ...870L..18Q}
{Qin}, Y., {Marchant}, P., {Fragos}, T., {Meynet}, G., \& {Kalogera}, V. 2019,
  \apjl, 870, L18

\bibitem[{{Rees}(1988)}]{rees88}
{Rees}, M.~J. 1988, \nat, 333, 523

\bibitem[{{Rhode} \& {Zepf}(2001)}]{rhode01}
{Rhode}, K.~L. \& {Zepf}, S.~E. 2001, \aj, 121, 210

\bibitem[{{Ripamonti} \& {Mapelli}(2012)}]{ripamonti12}
{Ripamonti}, E. \& {Mapelli}, M. 2012, \mnras, 423, 1144

\bibitem[{{Rizzuto} {et~al.}(2021{\natexlab{a}}){Rizzuto}, {Naab}, {Spurzem},
  {Giersz}, {Ostriker}, {Stone}, {Wang}, {Berczik}, \&
  {Rampp}}]{2021MNRAS.501.5257R}
{Rizzuto}, F.~P., {Naab}, T., {Spurzem}, R., {et~al.} 2021{\natexlab{a}},
  \mnras, 501, 5257

\bibitem[{{Rizzuto} {et~al.}(2021{\natexlab{b}}){Rizzuto}, {Naab}, {Spurzem},
  {Giersz}, {Ostriker}, {Stone}, {Wang}, {Berczik}, \& {Rampp}}]{rizzuto21}
{Rizzuto}, F.~P., {Naab}, T., {Spurzem}, R., {et~al.} 2021{\natexlab{b}},
  \mnras, 501, 5257

\bibitem[{{ROSAT Scientific Team}(2000)}]{1RXH}
{ROSAT Scientific Team}. 2000, VizieR Online Data Catalog, IX/28A

\bibitem[{{Saxton} {et~al.}(2020){Saxton}, {Komossa}, {Auchettl}, \&
  {Jonker}}]{saxton20}
{Saxton}, R., {Komossa}, S., {Auchettl}, K., \& {Jonker}, P.~G. 2020, \ssr,
  216, 85

\bibitem[{{Shen}(2019)}]{shen19}
{Shen}, R.-F. 2019, \apjl, 871, L17

\bibitem[{{Sheng} {et~al.}(2021){Sheng}, {Wang}, {Ferland}, {Shu}, {Yang},
  {Jiang}, \& {Chen}}]{sheng21}
{Sheng}, Z., {Wang}, T., {Ferland}, G., {et~al.} 2021, \apjl, 920, L25

\bibitem[{{Sivakoff} {et~al.}(2005){Sivakoff}, {Sarazin}, \&
  {Jord{\'a}n}}]{sivakoff05}
{Sivakoff}, G.~R., {Sarazin}, C.~L., \& {Jord{\'a}n}, A. 2005, \apjl, 624, L17

\bibitem[{{Sniegowska} {et~al.}(2020){Sniegowska}, {Czerny}, {Bon}, \&
  {Bon}}]{sniegowska20}
{Sniegowska}, M., {Czerny}, B., {Bon}, E., \& {Bon}, N. 2020, \aap, 641, A167

\bibitem[{{Stiele} \& {Kong}(2019)}]{stiele19}
{Stiele}, H. \& {Kong}, A.~K.~H. 2019, \apj, 877, 115

\bibitem[{{Str{\"u}der} {et~al.}(2001){Str{\"u}der}, {Briel}, {Dennerl},
  {Hartmann}, {Kendziorra}, {Meidinger}, {Pfeffermann}, {Reppin}, {Aschenbach},
  {Bornemann}, {Br{\"a}uninger}, {Burkert}, {Elender}, {Freyberg}, {Haberl},
  {Hartner}, {Heuschmann}, {Hippmann}, {Kastelic}, {Kemmer}, {Kettenring},
  {Kink}, {Krause}, {M{\"u}ller}, {Oppitz}, {Pietsch}, {Popp}, {Predehl},
  {Read}, {Stephan}, {St{\"o}tter}, {Tr{\"u}mper}, {Holl}, {Kemmer}, {Soltau},
  {St{\"o}tter}, {Weber}, {Weichert}, {von Zanthier}, {Carathanassis}, {Lutz},
  {Richter}, {Solc}, {B{\"o}ttcher}, {Kuster}, {Staubert}, {Abbey}, {Holland},
  {Turner}, {Balasini}, {Bignami}, {La Palombara}, {Villa}, {Buttler},
  {Gianini}, {Lain{\'e}}, {Lumb}, \& {Dhez}}]{struder01}
{Str{\"u}der}, L., {Briel}, U., {Dennerl}, K., {et~al.} 2001, \aap, 365, L18

\bibitem[{{Tully} {et~al.}(2008){Tully}, {Shaya}, {Karachentsev}, {Courtois},
  {Kocevski}, {Rizzi}, \& {Peel}}]{tully08}
{Tully}, R.~B., {Shaya}, E.~J., {Karachentsev}, I.~D., {et~al.} 2008, \apj,
  676, 184

\bibitem[{{Turner} {et~al.}(2001){Turner}, {Abbey}, {Arnaud}, {Balasini},
  {Barbera}, {Belsole}, {Bennie}, {Bernard}, {Bignami}, {Boer}, {Briel},
  {Butler}, {Cara}, {Chabaud}, {Cole}, {Collura}, {Conte}, {Cros}, {Denby},
  {Dhez}, {Di Coco}, {Dowson}, {Ferrando}, {Ghizzardi}, {Gianotti}, {Goodall},
  {Gretton}, {Griffiths}, {Hainaut}, {Hochedez}, {Holland}, {Jourdain},
  {Kendziorra}, {Lagostina}, {Laine}, {La Palombara}, {Lortholary}, {Lumb},
  {Marty}, {Molendi}, {Pigot}, {Poindron}, {Pounds}, {Reeves}, {Reppin},
  {Rothenflug}, {Salvetat}, {Sauvageot}, {Schmitt}, {Sembay}, {Short},
  {Spragg}, {Stephen}, {Str{\"u}der}, {Tiengo}, {Trifoglio}, {Tr{\"u}mper},
  {Vercellone}, {Vigroux}, {Villa}, {Ward}, {Whitehead}, \& {Zonca}}]{turner01}
{Turner}, M.~J.~L., {Abbey}, A., {Arnaud}, M., {et~al.} 2001, \aap, 365, L27

\bibitem[{{Wang} {et~al.}(2021){Wang}, {Perna}, \& {Armitage}}]{Wang21}
{Wang}, Y.-H., {Perna}, R., \& {Armitage}, P.~J. 2021, \mnras, 503, 6005

\bibitem[{{Wilms} {et~al.}(2000){Wilms}, {Allen}, \& {McCray}}]{wilms00}
{Wilms}, J., {Allen}, A., \& {McCray}, R. 2000, \apj, 542, 914

\bibitem[{{Zalamea} {et~al.}(2010){Zalamea}, {Menou}, \&
  {Beloborodov}}]{zalamea10}
{Zalamea}, I., {Menou}, K., \& {Beloborodov}, A.~M. 2010, \mnras, 409, L25

\bibitem[{{Zampieri} \& {Roberts}(2009)}]{zampieri09}
{Zampieri}, L. \& {Roberts}, T.~P. 2009, \mnras, 400, 677

\bibitem[{{Zepf} {et~al.}(2007){Zepf}, {Maccarone}, {Bergond}, {Kundu},
  {Rhode}, \& {Salzer}}]{zepf07}
{Zepf}, S.~E., {Maccarone}, T.~J., {Bergond}, G., {et~al.} 2007, \apjl, 669,
  L69

\bibitem[{{Zepf} {et~al.}(2008){Zepf}, {Stern}, {Maccarone}, {Kundu},
  {Kamionkowski}, {Rhode}, {Salzer}, {Ciardullo}, \& {Gronwall}}]{zepf08}
{Zepf}, S.~E., {Stern}, D., {Maccarone}, T.~J., {et~al.} 2008, \apjl, 683, L139

\bibitem[{{Zhao} {et~al.}(2021){Zhao}, {Wang}, {Zou}, {Wang}, \&
  {Dai}}]{zhao21}
{Zhao}, Z.~Y., {Wang}, Y.~Y., {Zou}, Y.~C., {Wang}, F.~Y., \& {Dai}, Z.~G.
  2021, arXiv e-prints, arXiv:2109.03471

\end{thebibliography}

\end{document}